\documentclass{article}

\usepackage{arxiv}

\usepackage[utf8]{inputenc} 
\usepackage[T1]{fontenc}    
\usepackage[colorlinks=true, allcolors=red]{hyperref}       
\usepackage{url}            
\usepackage{booktabs}       
\usepackage{nicefrac}       
\usepackage{microtype}      
\usepackage{doi}

\usepackage{graphicx}
\usepackage{comment}
\usepackage{amsfonts}
\usepackage{amsmath}
\usepackage{amssymb}
\usepackage{pifont}
\usepackage{bm,upgreek}
\usepackage{mathrsfs}
\usepackage{url}

\usepackage[linesnumbered, ruled, vlined]{algorithm2e}
\SetKwRepeat{Do}{do}{while}
\usepackage{pbox}
\usepackage{float}

\title{A cheat sheet for probability distributions of orientational data}

\date{} 					

\author{ {\hypersetup{urlcolor=black} \href{https://orcid.org/0000-0002-3914-6029}{\includegraphics[scale=0.06]{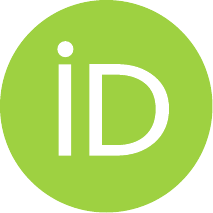}\hspace{1mm}P.C.~L{\'o}pez-Custodio}}\\
	Computer Science Department, Nottingham Trent University\\
	Clifton Campus, Clifton Ln \\
    Nottingham, NG11 8NS, United Kingdom\\
	\texttt{pablo.lopez-custodio@ntu.ac.uk}
}

\renewcommand{\shorttitle}{P.C. L{\'o}pez-Custodio, A cheat sheet for probability distributions of orientational data}

\hypersetup{
pdftitle={A cheat sheet for probability distributions of orientational data},
pdfsubject={engineering, statistics, orientations},
pdfauthor={P.C.~Lopez-Custodio},
pdfkeywords={},
}

\begin{document}
\maketitle

\begin{abstract}
The need for statistical models of orientations arises in many applications in engineering and computer science. Orientational data appear as sets of angles, unit vectors, rotation matrices or quaternions. In the field of directional statistics, a lot of advances have been made in modelling such types of data. However, only a few of these tools are used in engineering and computer science applications. Hence, this paper aims to serve as a cheat sheet for those probability distributions of orientations. Models for 1-DOF, 2-DOF and 3-DOF orientations are discussed. For each of them, expressions for the density function, fitting to data, and sampling are presented. The paper is written with a compromise between engineering and statistics in terms of notation and terminology. A Python library with functions for some of these models is provided. Using this library, two examples of applications to real data are presented. The code is available at {\small \url{https://github.com/PabloLopezCustodio/rotation-statistics}}.
\end{abstract}


\section{Introduction}

\noindent In engineering and computer science, we may often face the situation where statistics of a set of measurements of orientations are needed. These measurements might appear as a bunch of rotation matrices, quaternions or Euler angles. We know that, when our data are in $\mathbb{R}^d$, we can fit a multivariate normal distribution (MVN). However, we immediately notice that a direct application of the formula for the mean to our rotation matrices or quaternions results in neither a rotation matrix nor a quaternion. Hence, it is tempting to resort to the apparently unconstrained Euler angles. By looking carefully at our data we realise that two different triads of our Euler angles represent the same orientation. To avoid this redundancy, we force the first angle, $\alpha$, to be in $[0,\pi)$. After this fix, we realise that, if in several points of our dataset $\alpha$ is close to either 0 or $\pi$, the data will be unfairly split in two different sets that appear far away from each other. To make the situation even more ominous, we also remember the gimbal-lock problem. After trying to hack these issues, we realise that treating Euler angles as unconstrained points in $\mathbb{R}^3$ to fit a MVN is less than ideal. 

The scenario described above is studied in the field of directional statistics \cite{mardia_jupp}. Researchers in this field first noticed that measurements of wind directions could not be appropriately modelled using a normal distribution since the data lied in a circle, rather than in a straight line. The extension to higher dimensions followed naturally by studying cases of geological data, e.g. directions of long axes of pebbles from a glacial till \cite{arnold}, and remanent magnetism of lava flow \cite{fisher}, among many other problems. The advantage of these tools is that they were developed taking care of the manifold structure of the data. Specific models have been developed for angles (circular data), unit vectors (spherical data), rotation matrices (SO(3)) and quaternions (spherical data with antipodal symmetry).

In engineering and computer science, and more specifically in robotics, the need for statistics of orientations appears in a multitude of applications, for example, error propagation \cite{uncertainty_zhu,error_manipulator}, localisation \cite{banana, pose_estimation_book}, and end-effector pose estimation in continuum  \cite{continuum_robot_state_gp} and cable \cite{cable_robot_pose_estimation} robots. In robot-learning, the robot has to learn trajectories of both position and orientation of the end-effector when the demonstrations are captured in task space \cite{calinon_gmr_manifolds, caldwell_projection, orientational_promp,gmm_orientations, pc_nonparametric}. Related to the latter, many machine learning techniques need a statistical model of orientational or whole-pose data to operate. Direct examples of this are grasping \cite{grasp_arruda, grasp_lin, graspnet}, and orientation estimation by vision \cite{bingham_vision}.

In spite of these applications, there seems to be a lack of attention to the toolbox that directional statistics provides. Some common practices include fully treating the data as Euclidean and then forcing the results to lie on the corresponding manifold, or proceeding as explained above and hacking Euler-angles data. In other cases, when more care of the manifold structure is taken, the data are projected onto the tangent space of the manifold and a MVN is fitted there. Although this can give satisfactory results for very concentrated data, distortion may appear and the density function is not properly normalised. The lack of use of directional statistics tools might be due to the discrepancy in terminology and notation, and the belief that some of those models are too complex to manipulate or to run in real time.

This paper intends to serve as a cheat sheet for probabilistic models of orientational data, and provides the engineer and the computer scientist with the strictly necessary information that can be useful for our applications. For each model, we present its density function, how to fit it to data, and how to draw samples from it. These are the most commonly needed tasks for most applications. Although there are multiple methods for fitting and simulating the same model, we present the simplest techniques that give satisfactory results for engineering applications. This paper is accompanied by a Python library \cite{rotstats} that includes functions for several of these models. For the sake of clarity, the paper is written with a compromise in terminology and notation. The engineer will notice that we drop the hat from unit vectors since this will be used exclusively to denote the estimates of parameters. The statistician, however, will note that the term degrees of freedom (DOF) denotes here the number of variables needed to define the orientation of a rigid body, rather than the number of parameters of a model. To the knowledge of the author, no similar survey has been published. In the field of directional statistics, \cite{review_garcia_portugues} provides a survey on the state of the art but covers topics like regression, hypothesis testing and clustering, rather than presenting a list of models, and their simulation and fitting, which is the purpose of this paper.

This paper is organised as follows. In Section \ref{sec:background} some terminology and theoretical machinery required for the definition of those models is presented. The different models of probability distributions are then discussed. Those are split according to the DOFs of the orientational data. Models for 1-DOF, and 2-DOF orientations are presented in Sections \ref{sec:1dof} and \ref{sec:2dof}, respectively. For 3-DOF orientations we present models for quaternions in Section \ref{sec:3dof} and for rotation matrices in Section \ref{sec:SO3}. Section \ref{sec:diffusion} provides a brief introduction the use of diffusion equations to generate PDFs for orientational data. Examples with real data are presented in Section \ref{sec:experiments}. Finally, conclusions are provided in Section \ref{sec:conclusions}.


\section{Theoretical background and terminology}\label{sec:background}

\noindent In this paper, the {\it degrees of freedom} (DOF) refer to the number of variables needed to define the orientation of a rigid body with respect to another. Note that, unlike in statistics, the term is not used here to refer to the number of parameters of a model. In this paper, $d$ is the dimension of the ambient space. Hence, the set of unit vectors in $\mathbb{R}^d$ form the manifold $\mathcal{S}^{d-1}=\left\{\mathbf{x}\in\mathbb{R}^d\,|\,\|\mathbf{x}\|=1\right\}$. Note that we drop the hat from unit vectors and reserve it for the estimates of parameters, e.g. $\hat{\mu}$ is the estimate of $\mu$. Vectors are considered to be column vectors, i.e. $\mathbf{x}\in\mathbb{R}^d$ has shape $d\times 1$. In addition, in this paper the term {\it axis} refers to each of the three unit vectors conforming a right-handed coordinate system. This defers from its use in directional statistics, where an axis is a unit vector that is interchangeable by its antipodally symmetric vector.

\subsection{Parametrisations of orientations}\label{sec:param}

\begin{figure}[ht]
    \centering
    \includegraphics[width=0.55\textwidth]{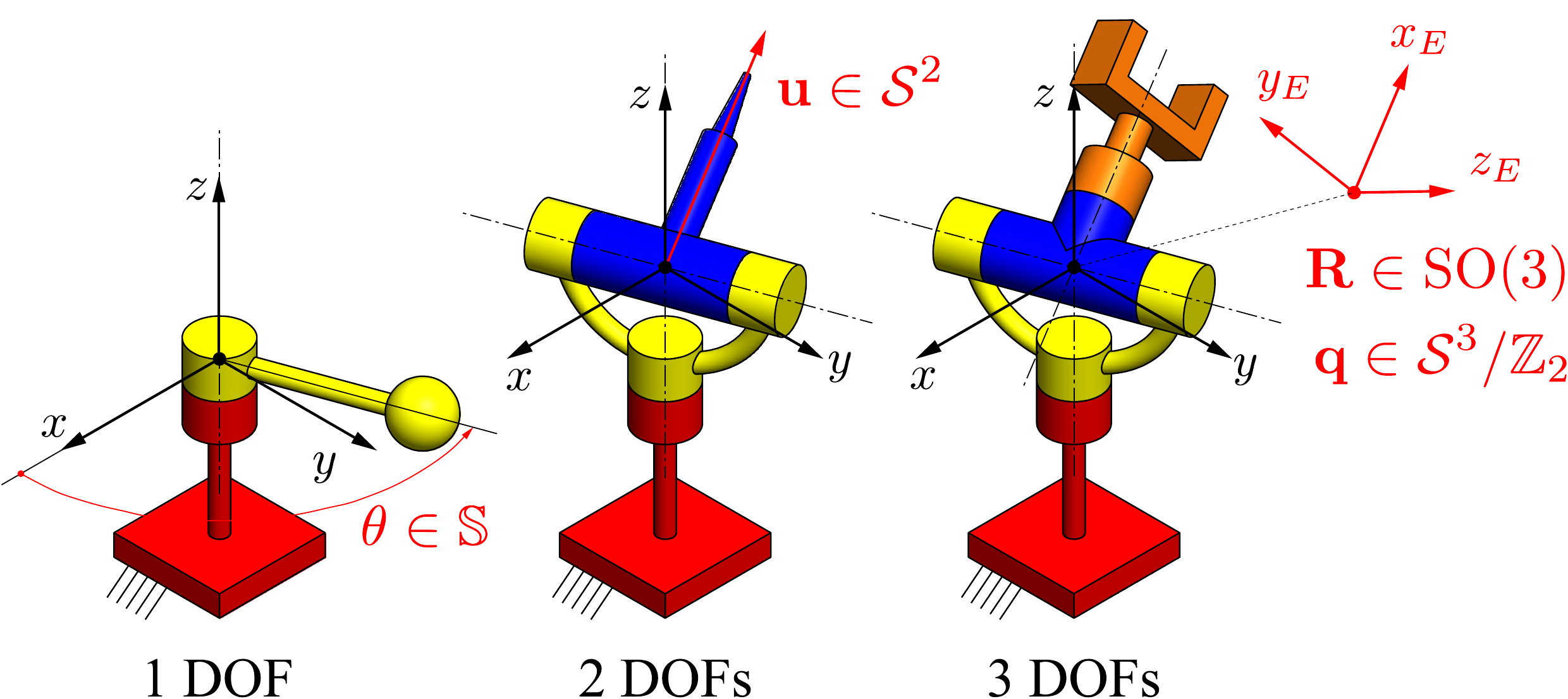}
    \caption{Parametrisation of orientations produced by 1-DOF, 2-DOF and 3-dOF rotations. For the 3-DOF case, the frame $E$ has origin at the centre of the spherical wrist but is rigidly attached to the end-effector.}
    \label{fig:dofs}
\end{figure}

The different models presented in this paper are divided into three sections according to the number of DOFs that the orientational data possess; see Fig. \ref{fig:dofs}.

{\bf 1-DOF orientations:} Consider the case of orientations produced by rotations about a fixed axis (Fig. \ref{fig:dofs}, left). These orientations form the manifold of a circle $\mathcal{S}^1$ and the group $R_1$. However, we will prefer to parametrise these orientations with the angle $\theta \in\mathbb{S}$, so that $(\cos \theta,\sin \theta)^{\top}\in\mathcal{S}^1$. In robotics and mechanisms theory, data of this type appears as the measurements of a single joint angle, the heading of a differential robot, the angle about the $Z-$axis in a pick-and-place task, and in general any data in the Schoenflies group of motion, $X_4$, where statistics of the orientational component are needed.

{\bf 2-DOF orientations:} Consider the orientation of a unit vector, for example by adjustment of elevation and azimuth angles (Fig. \ref{fig:dofs}, centre). In statistics, these are called {\it directions} in $\mathbb{R}^3$ and each is represented by a unit vector $\mathbf{u}\in \mathcal{S}^2$. Examples of data in $\mathcal{S}^2$ appear in the orientation of spinning tools as end-effectors of robots and CNC machines. Other examples include the tracking of cylindrical objects and normals to surfaces. This set of orientations does not form a group. 

{\bf 3-DOF orientations:} We represent the general 3-DOF orientation of a rigid body (Fig. \ref{fig:dofs}, right) with respect to an inertial frame using two different parametrisations. The first is the set of {\it rotation matrices} that form the special orthogonal group, SO(3), defined as:
\[
    \mathrm{SO}(3):=\left\{\mathbf{R}\in\mathbb{R}^{3\times 3}\,|\,\mathbf{R}\mathbf{R}^{\top}=\mathbf{R}^{\top}\mathbf{R}=\mathbf{I}_3,\,\det(\mathbf{R})=1\right\}
\]

The second representation is the {\it unit quaternion}. Consider the map $\mathrm{Rod}\,:\,\mathcal{S}^{3}\rightarrow\mathrm{SO}(3)$ defined by the Rodrigues' formula:
\begin{equation}\label{eq:rodrigues}    
\mathrm{Rod}(\mathbf{q}):=\mathbf{I}_3+2w\,\mathrm{skew}(x,y,z)+2\,\mathrm{skew}(x,y,z)^2
\end{equation}
\noindent where $\mathbf{q}:=(x,y,z,w)^{\top}\in\mathcal{S}^{3}$ is a quaternion, and 
\begin{equation}\label{eq:skew}
    \mathrm{skew}(x,y,z):=\left(\begin{array}{ccc}0& -z & y \\
         z& 0 & -x \\
         -y& x & 0\\
    \end{array}\right).
\end{equation}

Note that $\mathrm{Rod}(\mathbf{q})=\mathrm{Rod}(-\mathbf{q})$, $\forall \mathbf{q}\in\mathcal{S}^{3}$. Due to this antipodal redundancy, SO(3) is isomorphic to $\mathcal{S}^{3}/\mathbb{Z}_2\cong P\mathbb{R}^3$. $\mathcal{S}^{d-1}/\mathbb{Z}_2$ is the manifold where $\mathbf{x}\in\mathcal{S}^{d-1}$ represents the same element as $-\mathbf{x}$. Hence, we can represent a 3-DOF rotation as two antipodal points on a hypersphere of dimension 3 \--- these elements are called {\it axes} in directional statistics. The conversion from quaternion to rotation matrix requires a more detailed treatment due to numerical instability and several methods have been proposed. The reader is referred to \cite{mat_2_quat_survey} for a survey of those methods. 

In the case of quaternions, the composition operation is the quaternion product which in this paper is explicitly denoted by $*$ to avoid confusion with other products. Let $\mathbf{q}_i=(x_i,y_i,z_i,w_i)^{\top}\in\mathcal{S}^{3}/\mathbb{Z}_2$, $i=1,2$, be two quaternions and define $\mathbf{v}_i=(x_i,y_i,z_i)^{\top}$, $i=1,2$, then their product is $\mathbf{q}_3=\mathbf{q}_1*\mathbf{q}_2$$=(x_3,y_3,z_3,w_3)^{\top}$, where:
\[
w_3 = w_1w_2-\mathbf{v}_1^{\top}\mathbf{v}_2,\;\;\;\;\;\;(x_3,y_3,z_3)^{\top}=w_1\mathbf{v}_2+w_2\mathbf{v}_1+\mathbf{v}_1\times\mathbf{v}_2
\]
The inverse element of quaternion $\mathbf{q}=(x,y,z,w)$ is $\mathbf{q}^{-1}=(-x,-y,-z,w)$. 

\subsection{Descriptive measures for data in $\mathcal{S}^{d-1}$}\label{sec:descriptive}

Since directions and quaternions can be represented as points in $\mathcal{S}^2$ and $\mathcal{S}^3$, respectively, it is important to define the following quantities which are used in the models for data in $\mathcal{S}^{d-1}$. Consider the dataset $\mathscr{D}=\left\{\mathbf{x}_i\in\mathcal{S}^{d-1}\right\}_{i=1}^N$, then the {\it sample mean} is given by \cite{mardia_jupp}:

\begin{equation}\label{eq:sample_mean}
    \overline{\mathbf{x}}=\frac{1}{N}\sum^{N}_{i=1}\mathbf{x}_i = \overline{\rho}\overline{\mathbf{x}}_0,
\end{equation}

\noindent where $\overline{\rho}=\|\overline{\mathbf{x}}\|$ is the {\it mean resultant length}, and $\overline{\mathbf{x}}_0\in\mathcal{S}^2$, the {\it mean direction}. 

The {\it scatter matrix} or {\it inertia matrix} about the origin gives a measure of dispersion of $\mathscr{D}$:
\begin{equation}\label{eq:scatter}
    \mathbf{S}=\frac{1}{N}\sum^N_{i=1}\mathbf{x}_i\mathbf{x}_i^{\top}    
\end{equation}

\subsection{The acceptance-rejection method for sampling}\label{sec:accept_reject}

Several simulation methods included in this paper are based on a general technique known as {\it acceptance-rejection} \cite{kent_mardia_simulate}. Let $\mathcal{F}$ be a probability distribution that is difficult to simulate and has density function $f(x)$. The acceptance-rejection method allows to sample $\mathcal{F}$ if there is another distribution $\mathcal{G}$ (with density function $g(x)$) that is easy to simulate and the ratio $f(x)/g(x)$ is bounded by a constant. It is then said that $\mathcal{G}$ {\it envelops} $\mathcal{F}$. The big advantage of the acceptance-rejection method is that it does not require the computation of the normalising constant of $f(x)$. 

Let $f(x)=c_ff^{*}(x)$ and $g(x)=c_gg^{*}(x)$, where $c_f$ and $c_g$ are the corresponding normalising constants. Then, the method involves finding a constant $M^{*}$ such that $f^{*}(x)\leq M^{*}g^{*}(x)$, $\forall x$. Then, simulation is done following the scheme in algorithm \ref{alg:acc_rej}.

\begin{algorithm}
    \DontPrintSemicolon
    \KwIn{$f^{*}(x)$, $g^{*}(x)$, $M^{*}$}
    \KwOut{$Y\sim\mathcal{F}$}
    \While{\textnormal{\texttt{True}}}{
        $W \sim \mathrm{Unif}(0,1)$\;
        $X \sim \mathcal{G}$\;
        \If{$W < f^{*}(X)/(M^{*}g^{*}(X))$}{
            $Y \gets X$ \tcp*[f]{accept, otherwise reject}\; 
            \bf{break}\;
        }
    }
  \caption{The general acceptance-rejection scheme where $f^{*}(x)$ and $g^{*}(x)$ are the non-normalised density functions of models $\mathcal{F}$ and $\mathcal{G}$, respectively. Unif(0,1) is the uniform distribution in [0,1].}\label{alg:acc_rej}
\end{algorithm}

To evaluate the efficiency of the sampling, consider $M=c_fM^{*}/c_g\geq 1$. The number of iterations that algorithm \ref{alg:acc_rej} needs to accept a sample has mean $M$. Thus, the efficiency is $1/M$. However, note that $M$ does depend on the normalising constant $c_f$ which can be computationally expensive.

\subsection{Visualisation of distributions of orientational data}\label{sec:visualisation}

In the case of 3-DOF orientations, it is difficult to visualise the probability distributions due to the dimensions of the ambient spaces of $\mathcal{S}^{3}$ and SO(3). In this Section we follow the method to visualise distributions in SO(3) that was presented in \cite{lee_visual}. Let $\{\mathbf{e}_1,\mathbf{e}_2,\mathbf{e}_3\}$ be the canonical basis for $\mathbb{R}^3$. Then $\mathbf{R}\mathbf{e}_i\in\mathcal{S}^2$ is the $i$th axis of a frame with orientation defined by $\mathbf{R}\in\mathrm{SO}(3)$ with respect to the inertial frame. The method consists in plotting on $\mathcal{S}^2$, for each of the three axes, the probability of every unit vector being the $i$th axis of the rotated frame, namely for every $\mathbf{u}\in\mathcal{S}^2$, we plot $p(\mathbf{u}=\mathbf{R}\mathbf{e}_i)$. However, since there is an infinity of frames whose $i$th axes coincide with $\mathbf{u}$ \---- the constraint leaving 1 DOF to rotate about $\mathbf{u}$ \---- it is necessary to consider the density of all those rotation matrices to plot the density at $\mathbf{u}$. 

Hence, consider the set $H_i(\mathbf{u})=\{\mathbf{R}\in\mathrm{SO}(3)\,|\,\mathbf{R}\mathbf{e}_i=\mathbf{u}\}$, $i=1,2,3$. $H_i$ can be parametrised in terms of the redundant DOF that allows rotation about $\mathbf{u}$. However, we first need to define a ``home'' orientation. Let such orientation be the following \cite{lee_visual}:
\[
    \mathbf{R}_{O,i}(\mathbf{u}):=\exp\left[\frac{\arccos(\mathbf{e}_i\cdot\mathbf{u})}{\|\mathbf{e}_i\times\mathbf{u}\|}\mathbf{e}_i\times\mathbf{u}\right]
\]
\noindent where exp($\cdot$) is the exponential map of SO(3), see Eq. (\ref{eq:exp_log_so3}). Then, $H_i$ can be rewritten as $H_i(\mathbf{u}) = \{\mathbf{R}_{O,i}(\mathbf{u})\exp(\theta\mathbf{e}_i)\,|\,\theta\in\mathbb{S}\}$

Let $p_R(\cdot)$ be the density function on SO(3) that we want to visualise. Then we can marginalise the rotation matrices in the set $H_i$ and obtain the density at all $\mathbf{u}$ in $\mathcal{S}^2$ for the $i$th axis as follows:
\begin{equation}\label{eq:visual_R}
    p^{R}_i(\mathbf{u})=\frac{1}{2\pi}\int_{\theta\in\mathbb{S}}p_R\left(\mathbf{R}_{O,i}(\mathbf{u})\exp(\theta\mathbf{e}_i)\right)\mathrm{d}\theta,\;\; i=1,2,3.
\end{equation}

For distributions in $\mathcal{S}^3/\mathbb{Z}_2$, the method can be expressed in terms of quaternions by considering the quaternion of the home orientation:
\[
    \mathbf{q}_{O,i}(\mathbf{u}):=\cos\alpha + \sin\alpha\frac{\mathbf{e}_i\times\mathbf{u}}{\|\mathbf{e}_i\times\mathbf{u}\|},\;\;\alpha=\frac{\arccos(\mathbf{e}_i\cdot\mathbf{u})}{2}.
\]
Then, if the quaternion density function we want to visualise is $p_Q(\cdot)$,  it follows that:
\begin{equation}\label{eq:visual_q}
    p_i^Q(\mathbf{u})=\frac{1}{\pi}\int_{\theta\in\mathbb{S}}p_Q\left[\mathbf{q}_{O,i}(\mathbf{u})*\left(\cos\frac{\theta}{2}+\sin\frac{\theta}{2}\mathbf{e}_i\right)\right]\mathrm{d}\theta,\;\; i=1,2,3.
\end{equation}

Eqs (\ref{eq:visual_R}) (or (\ref{eq:visual_q})) provide a plot on $\mathcal{S}^2$ for each of the three axes of the frame. If the densities are concentrated, it is possible to merge the three plots to more clearly visualise the density of each axis around the mean orientation of the frame. Examples of this type of visualisations are shown in Figs. \ref{fig:mf}, \ref{fig:aruco} and \ref{fig:listerine_distributions}.


\section{1-DOF orientations: Angles, $\mathbb{S}$}\label{sec:1dof}

\noindent In this section, models for orientational data produced by 1-DOF rotations, as described in Fig. \ref{fig:dofs}, are presented. Hence, the data is in $\mathbb{S}$.


\subsection{The wrapped normal distribution}\label{sec:wrapped}

A family of distributions on the circle can be generated by wrapping a distribution for the line around a unit circle. Probably the most popular of such models is obtained by wrapping $\mathcal{N}(\mu, \sigma^2)$ around a circle. The result is the wrapped normal distribution, $\mathrm{WN}(\mu,\rho)$, which has density function for $x\in\mathbb{S}$ \cite{mardia_jupp}:
\begin{equation}\label{eq:f_wn}
    f_{\mathrm{WN}}(x\,|\,\mu,\rho):=\frac{1}{2\pi}\left\{1+2\sum^{\infty}_{k=1}\rho^{k^2}\cos \left(k(x-\mu)\right)\right\},
\end{equation}
\noindent where $\mu\in\mathbb{S}$ is the mean, and $\rho\in[0,1]$ is the concentration parameter which satisfies $\sigma^2=-2\log\rho$.The distribution is unimodal and is symmetric with respect to $\mu$. 

Other wrapped models include the wrapped Poisson Distribution and the Wrapped Cauchy Distribution. For details of this family of distributions see Section 3.5.7 of \cite{mardia_jupp}.

{\bf Parameter estimation:} Given a dataset $\mathscr{D}=\left\{x_i\in\mathbb{S}\right\}^{N}_{i=1}$, $\mu$ and $\rho$ can be estimated using the method of moments as follows \cite{WN_MLE}. 
\begin{equation}\label{eq:wn_mle}
    \hat{\mu} = \mathrm{atan2}\left[\mathbf{r}^{\top}(0,1), \mathbf{r}^{\top}(1,0)\right],\;\;\hat{\rho} = \|\mathbf{r}\|,\;\;\text{where}\;\;\mathbf{r}:=\frac{1}{N}\sum^N_{i=1}(\cos x_i, \sin x_i)^{\top}
\end{equation}

{\bf Simulation:} To draw samples from $\mathrm{WN}(\mu,\rho)$, simply simulate $y\sim\mathcal{N}(\mu,-2\log\rho)$, then $x\sim \mathrm{WN}(\mu,\rho)$ if $x=\mathrm{atan2}(\sin y, \cos y)\in(-\pi,\pi]$

\subsection{The von Mises distribution}\label{sec:vM}

Since the density function of the wrapped normal distribution involves an infinite series, the von Mises distribution, $\mathrm{vM}(\mu, \kappa)$, provides an approximation that overcomes this drawback at the cost of more complexity. Its density function at $x\in\mathbb{S}$ is given by \cite{mardia_jupp}:

\begin{equation}\label{eq:f_vM}
    f_{\mathrm{vM}}(x\,|\,\mu,\kappa):=\frac{1}{2\pi I_0(\kappa)}\mathrm{exp}\left(\kappa\cos(\theta-\mu)\right)
\end{equation}

\noindent where $\mu\in\mathbb{S}$ and $\kappa\in\mathbb{R}$ represent, respectively, the mean and the concentration parameter. $I_0$ is the modified Bessel function of the first kind of order 0, see Eq. (\ref{eq:I0_I1}).

\begin{figure}[ht]
    \centering
    \includegraphics[width=0.5\textwidth]{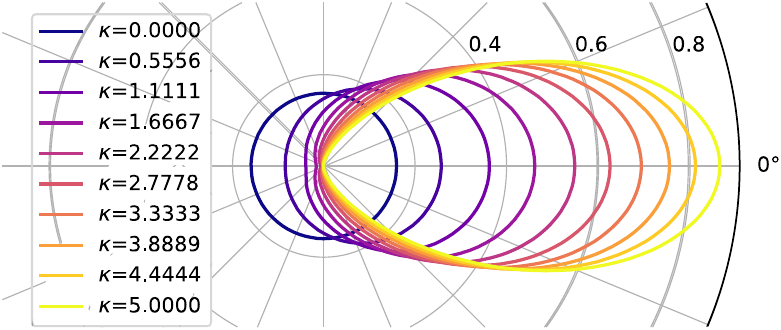}
    \caption{Density of the von Mises distribution for $x\in\mathbb{S}$ and ten equidistant values of $\kappa$ from 0 to 5.}
    \label{fig:vM}
\end{figure}

The distribution is unimodal and $f_{\mathrm{vM}}(x\,|\,\mu,\kappa)$ is symmetric with respect to $\mu$ reaching a minimum at $x=-\mu$. Due to the indeterminacy $\mathrm{vM}(\mu+\pi, \kappa)=\mathrm{vM}(\mu, -\kappa)$, normally it is considered that $\kappa\geq 0$. The larger the value of $\kappa$, the more concentrated the distribution. If $\kappa=0$ the distribution is uniform, whilst for $\kappa\rightarrow\infty$ the distribution approximates $\mathcal{N}(\mu,\kappa^{-1})$. Fig. \ref{fig:vM} shows the density $f_{\mathrm{vM}}(x\,|\,\mu,\kappa)$ for several values of $\kappa$ from 0 to 5.

A Python class for the vM distribution is available from \texttt{SciPy} as \texttt{vonmises} from the \texttt{stats} library.

{\bf Parameter estimation:} Given a dataset $\mathscr{D}=\left\{x_i\in\mathbb{S}\right\}^{N}_{i=1}$, the parameters of the von Mises distribution can be estimated by considering $d=2$ in the parameter fitting procedure presented in Section \ref{sec:vMF}. It follow that $\hat{\mu} = \mathrm{atan2}\left[\mathbf{r}^{\top}(0,1), \mathbf{r}^{\top}(1,0)\right]$ with $\mathbf{r}$ defined as in Eq. (\ref{eq:wn_mle}). By letting $\overline{\rho}=\|\mathbf{r}\|$ and $d=2$, $\kappa$ can be estimated by either Eqs (\ref{eq:vMF_hat_kappa_1}) or (\ref{eq:vMF_hat_kappa_2}).

{\bf Simulation:} Sampling from the von Mises distribution can be done by means of acceptance-rejection (Section \ref{sec:accept_reject}) using the wrapped Cauchy distribution as envelope \cite{mardia_jupp}. The resulting scheme is shown in Algorithm \ref{alg:vM_sim}.

\begin{algorithm}
    \DontPrintSemicolon
    \KwIn{$\mu$, $\kappa$}
    \KwOut{$x\sim\mathrm{vM}(\mu,\kappa)$}
    $a \gets 1+\sqrt{1+4\kappa^2}$\;
    $b \gets \left(a-\sqrt{2a}\right)/(2\kappa)$\;
    $r \gets (1+b^2)/(2b)$\;
    \While{\textnormal{\texttt{True}}}{
        $U_1, U_2, U_3 \sim \mathrm{Unif}(0,1)$\;
        $z \gets \cos(\pi U_1)$\;
        $f \gets (1+rz)/(r+z)$\;
        $c \gets \kappa(r-f)$\;
        \If{$c(2-c)-U_2>0$ \textnormal{\bf{or}} $\mathrm{log}(c/U_2)+1-c>0$}{
            $x \gets \mu + \mathrm{sign}(U_3-0.5)\cos^{-1}(f)$\;
            \bf{break}\;
        }
    }
  \caption{The acceptance-rejection algorithm for simulating the von Mises distribution}\label{alg:vM_sim}
\end{algorithm}


\section{2-DOF orientations: Directions, $\mathcal{S}^2$}\label{sec:2dof}

\noindent This section presents models for orientations produced by 2-DOF rotations (Fig. \ref{fig:dofs}, centre). Thus, the data are in $\mathcal{S}^2$ and the models do not have antipodal symmetry. However, since it is also possible to apply them to quaternions ($d=4$) after sign correction, the information is presented for general $\mathcal{S}^{d-1}$ when possible.

\subsection{The von Mises-Fisher distribution}\label{sec:vMF}

The von Mises Fisher distribution, vMF, is a $d-$dimensional extension of the vM distribution for $\mathcal{S}^1$ ($d=2$) from Sec. \ref{sec:vM}. This model is also known as the {\it Langevin distribution} as a generalisation due to Langevin \cite{langevin} was known since 1905 in magnetism theory. The form commonly used in directional statistics was popularised by Watson and Williams \cite{watson_vmf}.

The density function of vMF at $\mathbf{x}\in\mathcal{S}^{d-1}$ is given by \cite{watson_vmf}:
\begin{equation}\label{eq:f_vMF}
    f_{\mathrm{vMF}}(\mathbf{x}\,|\,\bm{\upmu},\kappa):=\frac{\kappa^{\frac{d}{2}-1}}{(2\pi)^{\frac{d}{2}}I_{\frac{d}{2}-1}(\kappa)}\mathrm{exp}\left(\kappa\bm{\upmu}^{\top}\mathbf{x}\right)
\end{equation}

\noindent where $\kappa>0$ is the concentration parameter, $\bm{\upmu}\in\mathcal{S}^{d-1}$ is the mean, and $I_{d/2-1}(\kappa)$ is the modified Bessel function of the first kind and order $\nu=d/2-1$ as defined in Eq. (\ref{eq:Iv}). The special case on $\mathcal{S}^2$ ($d=3$) is known as {\it Fisher distribution} due to the seminal paper by Fisher \cite{fisher}. In this case, Eq. (\ref{eq:f_vMF}) simplifies to \cite{fisher}:
\begin{equation}
    f_{\mathrm{vMF}}(\mathbf{x}\,|\,\bm{\upmu},\kappa):=\frac{\kappa}{\sinh\kappa}\mathrm{exp}\left(\kappa\bm{\upmu}^{\top}\mathbf{x}\right)
\end{equation}

\begin{figure}[ht]
    \centering
    \includegraphics[width=0.7\textwidth]{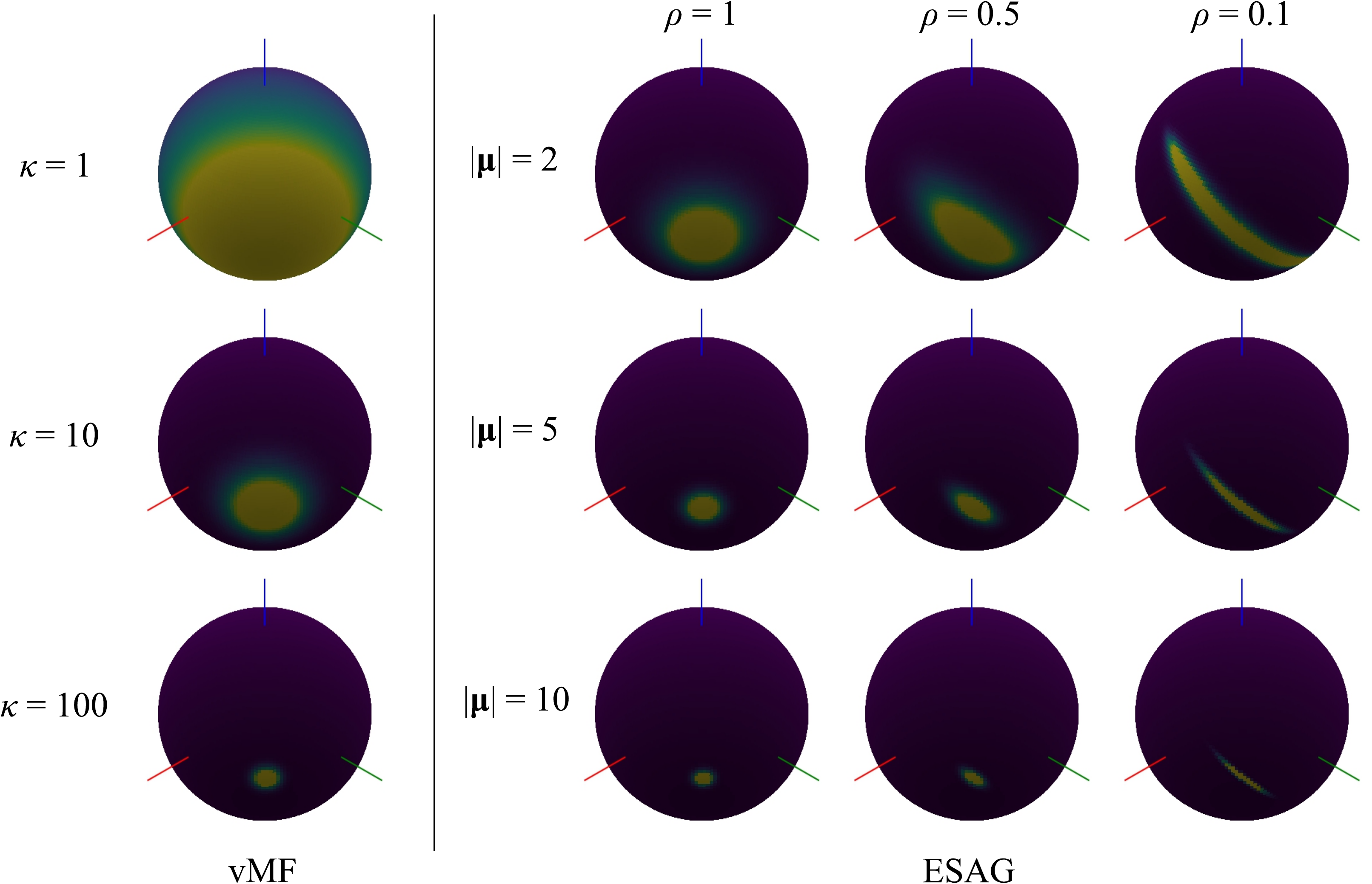}
    \caption{Effect of concentration parameters. Let $\mathbf{R}=\exp\{(0,0,\pi/4)^{\top}\}$. LEFT: vMF with $\bm{\upmu}=\mathbf{R}(1,0,0)^{\top}$. RIGHT: ESAG with $\bm{\upxi}_3=\mathbf{R}(1,0,0)^{\top}$, $\bm{\upxi}_2=\mathbf{R}(0, -\sqrt{2}/2, \sqrt{2}/2)^{\top}$, $\bm{\upxi}_1=\bm{\upxi}_2\times\bm{\upxi}_3$. This leads to $\psi=\pi/4$}.
    \label{fig:vMF_ESAG}
\end{figure}

Note that $\kappa$ is the only parameter modeling the concentration. Hence, vMF is symmetric around the mean $\bm{\upmu}$, which results in circular density contours, i.e. the distribution is {\it isotropic}. In addition, note that for $d=2$, vMF is reduced to the vF for circular data presented in Section (\ref{sec:vM}). Fig. \ref{fig:vMF_ESAG} shows the effect of $\kappa$ on the distribution.

A Python class for the vMF distribution is available from \texttt{SciPy} as \texttt{vonmises\textunderscore fisher} from the \texttt{stats} library.

{\bf Parameter estimation:} Given a dataset $\mathscr{D}=\left\{\mathbf{x}_i\in\mathcal{S}^{d-1}\right\}^{N}_{i=1}$, the mean is estimated as the mean direction defined in Eq. (\ref{eq:sample_mean}), $\hat{\bm{\upmu}}=\overline{\mathbf{x}}_0$.

The maximum likelihood estimate (MLE) of $\kappa$ satisfies:

\begin{equation}\label{eq:vM_MLE_kappa}
    \frac{I_{\frac{d}{2}}(\hat{\kappa})}{I_{\frac{d}{2}-1}(\hat{\kappa})}=\overline{\rho},
\end{equation}

\noindent where $\overline{\rho}$ is the mean resultant length defined in Eq. (\ref{eq:sample_mean}). If $\kappa\gg2$, an approximate solution for Eq. (\ref{eq:vM_MLE_kappa}) is proposed in Eqs. (10.3.7) and (10.3.10) of  \cite{mardia_jupp}:
\begin{equation}\label{eq:vMF_hat_kappa_1}
    \begin{array}{cc}
        \displaystyle \hat{\kappa} \approx \frac{d-1}{d(1-\overline{\rho})} & \mathrm{for\;\;large} \;\;\overline{\rho} \\
        \displaystyle \hat{\kappa} \approx d\overline{\rho}\left(1+\frac{d}{d+2}\overline{\rho}^2+\frac{d^2(d+8)}{(d+2)^2(d+4)}\overline{\rho}^4\right) & \mathrm{for\;\;small}\;\; \overline{\rho}
    \end{array}
\end{equation}

More precise approximations involve the computation of Bessel functions \cite{vMF_shortnote}. For example, in \cite{tanabe_vMF_mle}, the following approximation is proposed:
\begin{equation}\label{eq:vMF_hat_kappa_2}
    \hat{\kappa} = \frac{\kappa_l\Phi(\kappa_u)-\kappa_u\Phi(\kappa_l)}{\Phi(\kappa_u)-\Phi(\kappa_l)-\kappa_u+\kappa_l}
\end{equation}
\noindent where:
\begin{equation}
    \kappa_l=\frac{\overline{\rho}(d-2)}{1-\overline{\rho}^2},\;\kappa_u=\frac{d\overline{\rho}}{1-\overline{\rho}^2},\;\Phi(\kappa)=\overline{\rho}\kappa\frac{I_{\frac{d}{2}-1}(\kappa)}{I_{\frac{d}{2}}(\kappa)}
\end{equation}

The estimation is bound by $\kappa_l$ and $\kappa_u$, i.e. $\kappa_l\leq\hat{\kappa}\leq\kappa_u$. A systematic benchmarking of this and other methods to approximate $\hat{\kappa}$ is presented in \cite{vM_kappa_benchmark}

{\bf Simulation:} An efficient method to simulate vMF was proposed in \cite{vMF_sim_Wood}. Although the method is presented for $d$, the simpler case of $d=3$ is explained here. Let $\mathbf{R}\in\mathrm{SO(3)}$ be any rotation matrix that aligns the mean with the $Z-$axis, i.e. $\bm{\upmu}'=\mathbf{R}\bm{\upmu}=(0,0,1)^{\top}$. Note that $\mathbf{R}$ is not unique. Consider the polar parametrization, $\mathbf{x}'=(\cos\phi\sin\theta, \cos\phi\cos\theta, \sin\phi)\in\mathcal{S}^2$. Under this setting, simulate $\theta\sim\mathrm{Unif}(0,2\pi)$ and, to obtain $\phi$, draw $\xi\sim\mathrm{\mathrm{Unif}}(0,1)$ so that:
\[   
\phi = 
     \begin{cases}
       \arccos\left\{1+\frac{1}{\kappa}\left[\ln\xi+\ln\left(1-\frac{\xi-1}{\xi}e^{-2\kappa}\right)\right]\right\} &\quad\text{if } \xi\neq 0,\\
       \frac{\pi}{2} &\quad\text{if } \xi= 0.
     \end{cases}
\]
Then return the sample to the original frame, i.e. $\mathbf{x}=\mathbf{R}^{\top}\mathbf{x}'$.

\subsection{The Kent distribution}\label{sec:kent}

To extend the vMF distribution to a model with non-circular contours, a Fisher-Bingham distribution, $\mathrm{FB}_8$, was proposed which uses a total of 8 parameters. Kent \cite{kent} proposed to impose constraints on these parameters to simplify the model while preserving its non-circular level curves. The result is a 5-parameter model known as the Kent distribution, $\mathrm{FB}_5(\mathbf{Q},\kappa,\beta)$, with the following density function for $\mathbf{x}\in\mathcal{S}^2$ \cite{kent}:

\begin{equation}\label{eq:f_kent}
    f_{\mathrm{FB}_5}(\mathbf{x}\,|\,\mathbf{Q},\kappa,\beta):=\frac{1}{c(\kappa,\beta)}\mathrm{exp}\left\{\kappa\bm{\upgamma}^{\top}_1\mathbf{x}+\beta\left[(\bm{\upgamma}^{\top}_2\mathbf{x})^2-(\bm{\upgamma}^{\top}_3\mathbf{x})^2\right]\right\}
\end{equation}

\noindent where $\mathbf{Q}:=[\bm{\upgamma}_1,\bm{\upgamma}_2,\bm{\upgamma}_3]\in\mathrm{O}(3)$, so $\left\{\bm{\upgamma}_1,\bm{\upgamma}_2,\bm{\upgamma}_3\right\}$ are orthogonal unit vectors. $\bm{\upgamma}_1$ is the mean direction, and $\bm{\upgamma}_2$ and $\bm{\upgamma}_3$ are, respectively, the directions of the minor and major axes of the elliptical shape of the distribution. $\kappa\in\mathbb{R}$ is the concentration parameter (the larger $\kappa$, the more concentrated the distribution), and $\beta\in[0,\kappa/2)$ indicates how flat the elliptical shape is starting at $\beta=0$ where the distribution is circular and equal to the vMF distribution.

The normalising term is given by \cite{kent}:
\begin{equation}\label{c_kent}
    c(\kappa,\beta):= 2\pi\sum^{\infty}_{j=0}\frac{\Gamma(j+\frac{1}{2})}{j!}\beta^{2j}\left(\frac{2}{\kappa}\right)^{2j+\frac{1}{2}}I_{2j+\frac{1}{2}}(\kappa)
\end{equation}

Note that $c(\kappa,\beta)$ depends on the modified Bessel function of the first kind of order $2j+1/2$. Expressions for these functions are provided in Appendix \ref{sec:appendix_expressions}.

{\bf Parameter estimation:} Given a data set $\mathscr{D}=\{\mathbf{x}_i\in\mathcal{S}^2\}^N_{i=1}$, the parameters $\mathbf{Q}$, $\kappa$ and $\beta$ are more easily obtained by reparametrising the model as follows. Let $O$ be the world frame with respect to which the data is defined, and let its axes be the canonical basis $\{\mathbf{i}_O,\mathbf{j}_O,\mathbf{k}_O\}$. Note that we drop the hat from the usual notation of this basis to avoid confusion with the estimates of parameters. Define a frame $K$ with axes $\{\mathbf{i}_K,\mathbf{j}_K,\mathbf{k}_K\}$ such that:
\begin{equation}
        \mathbf{i}_K = \bm{\upgamma}_1,\;\;
        \mathbf{k}_K = \frac{\mathbf{i}_O\times\bm{\upgamma}_1}{\|\mathbf{i}_O\times\bm{\upgamma}_1\|},\;\;
        \mathbf{j}_K = \mathbf{k}_K \times \mathbf{j}_K.
\end{equation}

Since $\mathbf{i}_K = \bm{\upgamma}_1$, there is a $\psi$ such that:
\begin{equation}\label{eq:reparam_basis}
    \begin{split}
        \bm{\upgamma}_2 &=\mathrm{exp}(\psi\mathbf{k}_K)\mathbf{j}_K=\cos\psi\mathbf{j}_K + \sin\psi\mathbf{k}_K,\\
        \bm{\upgamma}_3 &=\mathrm{exp}(\psi\mathbf{k}_K)\mathbf{k}_K=-\sin\psi\mathbf{j}_K + \cos\psi\mathbf{k}_K,\\
    \end{split}
\end{equation}

\begin{figure}[ht]
    \centering
    \includegraphics[width=0.5\textwidth]{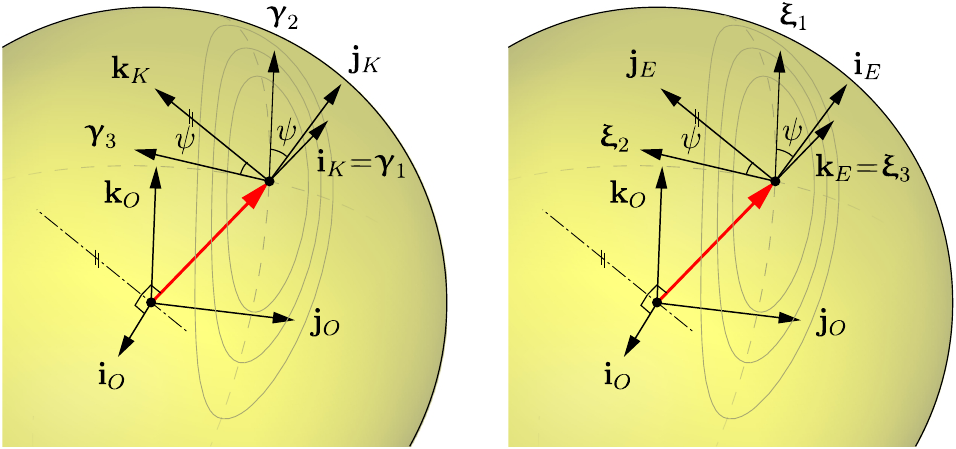}
    \caption{Frames used in the reparametrisation of $\mathrm{FB}_5$ (left) and ESAG (right) distributions (schematic).}
    \label{fig:esag_kent_frames}
\end{figure}

Fig. \ref{fig:esag_kent_frames} shows the frames involved in this setting. This a reparametrisation of the Kent distribution as $\mathrm{FB}_5(\kappa,\beta,\psi,\bm{\upgamma}_1)$. 

Those parameters can be estimated using the method of moments \cite{kent,kent_kasarapu}. The mean is given by the mean direction in Eq (\ref{eq:sample_mean}). Hence $\hat{\bm{\upgamma}}_1=\overline{\mathbf{x}}_0$. 

To estimate $\psi$, represent the data in $\mathscr{D}$ in frame $E$. Then the scatter matrix (defined in Eq. (\ref{eq:scatter})) is transformed to $\mathbf{B}:=({}^{O}_{E}\mathbf{R}^{\top})\mathbf{S}({}^{O}_{E}\mathbf{R})$, where ${}^{O}_{E}\mathbf{R}=[\mathbf{i}_E,\mathbf{j}_E,\mathbf{k}_E]$. Since $\hat{\bm{\upgamma}}_2$ and $\hat{\bm{\upgamma}}_3$ are the principal directions of dispersion, $\psi$ can be found by computing the eigendecomposition of the upper-left $2\times2$ submatrix of $\mathbf{B}$. Let $\mathbf{B}_L=[b_{ij}]_{i=1,j=1}^{2,2}$ be such submatrix, then \cite{kent_kasarapu}:
\[
    \hat{\psi} = \frac{1}{2}\arctan\left[\frac{2b_{12}}{b_{11}-b_{22}}\right],
\]
\noindent and $\hat{\bm{\upgamma}}_2$ and $\hat{\bm{\upgamma}}_3$ are obtained from Eq. (\ref{eq:reparam_basis}) or, equivalently, $\hat{\bf{Q}}=\mathrm{exp}(\hat{\psi}\hat{\bm{\upgamma}}_1)({}^{O}_{E}\mathbf{R})$.

To estimate the concentration and shape parameters, let $l_1>l_2$ be the eigenvalues of $\mathbf{B}_L$, and define $r_1:=\overline{\rho}$ and $r_2:=l_1-l_2$, then the following asymptotic approximations can be obtained:
\begin{equation}
    \begin{split}
        \hat{\kappa} &\approx (2-2r_1-r_2)^{-1} + (2-2r_1+r_2)^{-1} \\
        \hat{\beta} &\approx \frac{1}{2}\left[(2-2r_1-r_2)^{-1} - (2-2r_1+r_2)^{-1}\right]
    \end{split}
\end{equation}

{\bf Simulation:} Probably the simplest way to simulate $\mathrm{FB}_5$ is by means of an acceptance-rejection scheme with a uniform distribution on the sphere, $\mathrm{US}_2$, as envelope \cite{kents_student}. For this purpose it is easier to work in polar coordinates, i.e. $\mathbf{x}=(\sin\theta\cos\phi,\,\sin\theta\sin\phi,\,\cos\theta)\in\mathcal{S}^2$. With this parametrization, sampling from $\mathrm{US}_2$ is as simple as drawing $\phi\sim\mathrm{Unif}(0,2\pi)$ and $\cos\theta\sim\mathrm{Unif}(-1,1)$. Under this change of variable, $\mathrm{FB}_5$ has the following non-normalized density function:
\[
    f^{*}(\theta,\phi)=\mathrm{exp}\left\{\kappa\cos\theta+\beta\sin^2\theta\cos(2\phi)\right\}\sin\theta
\]
\noindent while the non-normalised density function of $\mathrm{US}_2$ is simply $g^{*}(\theta,\phi)=\sin\theta$. It follows that:
\begin{equation}\label{eq:kent_fstar}
    \frac{f^{*}(\theta,\phi)}{g^{*}(\theta,\phi)}=\exp\left\{\kappa\cos\theta+\beta\sin^2\theta\cos(2\phi)\right\}\;\;\mathrm{and}\;\;M^{*}=\exp(\kappa+\beta)
\end{equation}

Hence, $\mathrm{FB}_5$ can be simulated following Algorithm (\ref{sec:accept_reject}) with $\mathcal{G}$ as $\mathrm{US}_2$, and taking $f^{*}(X)/g^{*}(X)$ and $M^{*}$ from Eq. (\ref{eq:kent_fstar}).

\subsection{The elliptically symmetric angular Gaussian distribution}\label{sec:esag}

In a similar way that the Kent distribution simplifies the Fisher-Bingham distribution by imposing constraints while maintaining elliptical contours, it is possible to prescribe constraints on the parameters of the angular Gaussian distribution, sometimes referred to as projected Gaussian \cite{mardia_jupp}. The result of such simplification is a 5-parameter model with elliptical contours adequately named the elliptically symmetric Gaussian distribution (ESAG) \cite{esag}. 

The density of the ESAG distribution at $\mathbf{x}\in\mathcal{S}^2$ is given by \cite{esag}:

\begin{equation}\label{eq:esag}
    f_{\mathrm{ESAG}}(\mathbf{x}\,|\,\bm{\upmu},\,\mathbf{V}):=\frac{1}{2\pi(\mathbf{x}^{\top}\mathbf{V}^{-1}\mathbf{x})}
    \mathrm{exp}\Bigg[\frac{1}{2}\Bigg\{\frac{(\mathbf{x}^{\top}\bm{\upmu})^2}{\mathbf{x}^{\top}\mathbf{V}^{-1}\mathbf{x}}
    -\bm{\upmu}^{\top}\bm{\upmu}\Bigg\}\Bigg]
    M_2\left(\frac{\mathbf{x}^{\top}\bm{\upmu}}{(\mathbf{x}^{\top}\mathbf{V}^{-1}\mathbf{x})^{1/2}}\right),
\end{equation}

\noindent where:
\begin{eqnarray}
    M_2(\alpha)&:=&(1+\alpha^2)\Phi(\alpha)+\alpha\phi(\alpha),
\end{eqnarray}
\noindent where $\phi(\cdot)$ and $\Phi(\cdot)$ are the functions for standard normal probability density and cumulative density, respectively. For a treatment with any dimension $d$, see \cite{esag}.

The direction $\bm{\upmu}/\|\bm{\upmu}\|$ is the mean, while $\|\bm{\upmu}\|$ controls how concentrated the distribution is. The parameter $\mathbf{V}\in\mathrm{Sym}^{+}(3)$ has eigenvalues $\rho_i$ with corresponding eigenvectors $\bm{\upxi}_i$, $i=1,2,3$. The constraints $|\mathbf{V}|=1$ and $\mathbf{V}\bm{\upmu}=\bm{\upmu}$ are imposed. The latter has as consequence that, if $\bm{\upxi}_3 = \bm{\upmu}/\|\bm{\upmu}\|$, then $\rho_3=1$, while the former results in $\rho_1\rho_2\rho_3=1$. The principal axes of the elliptical contours are parallel to $\bm{\upxi}_1$ and $\bm{\upxi}_2$ and their shape is controlled by the proportion of $\rho_1$ and $\rho_2$. For convention, let $\rho_1\leq\rho_2$. Fig. \ref{fig:vMF_ESAG} shows the effect of $\rho$ and $\|\bm{\upmu}\|$ on a distribution with given $\bm{\upxi}_3$ and $\psi$.

{\bf Parameter estimation:} Similarly to the way we proceeded for $\mathrm{FB}_5$ in Section \ref{sec:kent}, for $d=3$ it is easier to reparametrise ESAG. Let $O$ be the world frame with respect to which the data is defined, and let its axes be the canonical basis $\{\mathbf{i}_O,\mathbf{j}_O,\mathbf{k}_O\}$. Define a frame $K$ with axes $\{\mathbf{i}_E,\mathbf{j}_E,\mathbf{k}_E\}$ such that:
\begin{equation}
        \mathbf{k}_E = \bm{\upxi}_3,\;\;
        \mathbf{j}_E = \frac{\mathbf{i}_O\times\bm{\upxi}_3}{\|\mathbf{i}_O\times\bm{\upxi}_3\|},\;\;
        \mathbf{i}_E = \mathbf{k}_E \times \mathbf{j}_E.
\end{equation}

Since $\mathbf{k}_E = \bm{\upgamma}_1$, there is a $\psi$ such that:
\begin{equation}
    \begin{split}
        \bm{\upxi}_1 &=\mathrm{exp}(\psi\mathbf{k}_E)\mathbf{i}_E=\cos\psi\mathbf{i}_E + \sin\psi\mathbf{j}_E,\\
        \bm{\upxi}_2 &=\mathrm{exp}(\psi\mathbf{k}_E)\mathbf{j}_E=-\sin\psi\mathbf{i}_E + \cos\psi\mathbf{j}_E,\\
    \end{split}
\end{equation}

In addition, let $\rho_1=\rho$ and $\rho_2=1/\rho$. Then the model is defined by $\bm{\upmu}\in\mathbb{R}^3$, $\psi\in\mathbb{S}$ and $\rho\in(0,1]$. To avoid working with restricted parameters, define:
\[
    \gamma_1 := \frac{1}{2}(\rho^{-1}-\rho)\cos(2\psi),\;\;
    \gamma_2 := \frac{1}{2}(\rho^{-1}-\rho)\sin(2\psi)
\]
\noindent then the inverse of the concentration matrix n Eq (\ref{eq:esag}) is given by:
\begin{equation}\label{eq:esag_V}
\begin{split}
    \mathbf{V}^{-1}=&\mathbf{I}_3+\gamma_1\left(\mathbf{i}_E\mathbf{i}_E^{\top}-\mathbf{j}_E\mathbf{j}_E^{\top}\right)+\gamma_2\left(\mathbf{i}_E\mathbf{j}_E^{\top}+\mathbf{j}_E\mathbf{i}_E^{\top}\right)
    \\
    &+\left[(\gamma_1^2+\gamma_2^2+1)^{1/2}-1\right]\left(\mathbf{i}_E\mathbf{i}_E^{\top}+\mathbf{j}_E\mathbf{j}_E^{\top}\right)
\end{split}
\end{equation}

Since $\bm{\upmu}$, $\gamma_1$ and $\gamma_2$ are unbounded, an optimization package can be used to obtain the parameters that maximise the likelihood for a dataset $\mathscr{D}=\{\mathbf{x}_i\in\mathcal{S}^2\}_{i=1}^N$.

\begin{equation}\label{eq:mle_esag}
    (\hat{\bm{\upmu}}, \hat{\gamma}_1, \hat{\gamma}_2) = \arg \max_{(\bm{\upmu}, \gamma_1, \gamma_2)} \sum_{i=1}^N\log f_{\mathrm{ESAG}}(\mathbf{x}_i\,|\, \bm{\upmu},\, \gamma_1, \gamma_2).
\end{equation}

\noindent note that, since the normalising term of ESAG does not depend on Bessel functions and has a simple closed form, (\ref{eq:mle_esag}) is easy to solve.

{\bf Simulation:} The biggest advantage of the ESAG distribution is how easy it is to simulate. To draw samples from $\mathrm{ESAG}(\bm{\upmu}, \mathbf{V})$, where $\mathbf{V}$ is defined in terms of $\gamma_1$, $\gamma_2$ and $\bm{\upmu}$ in Eq (\ref{eq:esag_V}), simulate $\mathbf{y}\sim\mathcal{N}(\bm{\upmu},\mathbf{V})$, then $\mathbf{x}\sim\mathrm{ESAG}(\bm{\upmu}, \mathbf{V})$ if $\mathbf{x}=\mathbf{y}/\|\mathbf{y}\|$.


\section{3-DOF orientations: Quaternions, $\mathcal{S}^3/\mathbb{Z}_2$}\label{sec:3dof}

\noindent This section includes models that have the desired feature of antipodal symmetry for quaternion data. However, since they can be used for directions too, i.e. $d=3$, when possible the information is presented for general $\mathcal{S}^{d-1}/\mathbb{Z}_2$.

\subsection{The Bingham distribution}\label{sec:bingham}

The Bingham distribution, B, is a model with antipodal symmetry and elliptical contours \cite{bingham}. For $\mathbf{x}\in\mathcal{S}^{d-1}/\mathbb{Z}_2$, two equivalent representations of the density function are commonly used depending on the way the paramters are treated \cite{bingham, glover_phd}:
\begin{eqnarray}\label{eq:bingham}
    f_B(\mathbf{x}\,|\,\mathbf{B})&:=&\frac{1}{F}\exp\left(\mathbf{x}^{\top}\mathbf{B}\mathbf{x}\right),\nonumber\\
    f_B(\mathbf{x}\,|\,\{\kappa_i,\mathbf{v}_i\}^{d}_{i=1})&:=&\frac{1}{F}\exp\left\{\sum^d_{i=1}\kappa_i\left(\mathbf{v}^{\top}_i\mathbf{x}\right)^2\right\},
\end{eqnarray}
\noindent where $\mathbf{B}\in\mathrm{Sym}^{+}(d)$ has eigenvalues $\kappa_1'\leq\ldots\leq \kappa_d'$ with corresponding eigenvectors $\mathbf{v}_1,\ldots,\mathbf{v}_d$. Since $f_B(\mathbf{x}\,|\,\{\kappa_i,\mathbf{v}_i\}^{d}_{i=1})=f_B(\mathbf{x}\,|\,\{\kappa_i+c,\mathbf{v}_i\}^{d}_{i=1})$, by convention we choose $\kappa_i = \kappa_i'-\kappa_d$ so that $\kappa_d=0$. 

The model has the desired antipodal symmetry since $f_B(\mathbf{x}\,|\,\mathbf{B})=f_B(-\mathbf{x}\,|\,\mathbf{B})$. The Bingham distribution has mean $\mathbf{v}_d$. Like in the models seen in Section \ref{sec:2dof}, the remaining eigenvectors are the principal directions of dispersion and their eigenvalues control the dispersion in each direction. 

The normalising constant $F$ depends exclusively on $\{\kappa_i\}^{d-1}_{i=1}$. In the case of quaternions ($d=4$) and directions ($d=3$), $F$ takes the respective forms:
\begin{eqnarray}\label{eq:bingham_c}
    F(\kappa_1,\kappa_2,\kappa_3)&=&2\sqrt{\pi}\sum^{\infty}_{i=0}\sum^{\infty}_{j=0}\sum^{\infty}_{k=0}\frac{\Gamma(i+\frac{1}{2})\Gamma(j+\frac{1}{2})\Gamma(k+\frac{1}{2})\kappa_1^{i}\kappa_2^{j}\kappa_3^{k}}{\Gamma(i+j+k+2)i!j!k!}\nonumber\\
    F(\kappa_1,\kappa_2)&=&2\sqrt{\pi}\sum^{\infty}_{i=0}\sum^{\infty}_{j=0}\frac{\Gamma(i+\frac{1}{2})\Gamma(j+\frac{1}{2})\kappa_1^{i}\kappa_2^{j}}{\Gamma(i+j+\frac{3}{2})i!j!}
\end{eqnarray}

Since $F$ is computationally expensive, methods to approximate it have been proposed,  \cite{kume_approx_normalising}.

{\bf Parameter estimation:} Given a dataset $\mathscr{D}=\{\mathbf{x}_i\in\mathcal{S}^{d-1}/\mathbb{Z}_2\}_{i=1}^N$, its scatter matrix $\mathbf{S}$ (see Eq. (\ref{eq:scatter})) is all that is needed to fit a Bingham distribution. Let $\lambda_1\geq\ldots\geq\lambda_d$ be the eigenvalues of $\mathbf{S}$, then the estimates $\{\hat{\mathbf{v}}_i\}_{i=1}^d$ are given by the corresponding eigenvectors of $\mathbf{S}$. Hence, the estimated mean, $\hat{\mathbf{v}}_d$, is the last eigenvector of $\mathbf{S}$.

The estimation of the concentration parameters $\kappa_i$ is more difficult. The MLEs of those parameters satisfy:
\begin{equation}\label{eq:bingham_mle}
    \frac{1}{F(\kappa_1,\ldots,\kappa_{d-1})}\cdot\frac{\partial F(\kappa_1,\ldots,\kappa_{d-1})}{\partial \kappa_i}=\lambda_i
\end{equation}
\noindent where $\partial F(\kappa_1,\ldots,\kappa_{d-1})/\partial\kappa_i$ is easy to compute from Eq. (\ref{eq:bingham_c}). Note that in the case of quaternions, Eq (\ref{eq:bingham_mle}) is a system of 3 equations in the unknowns $\{\kappa_1,\kappa_2,\kappa_3\}$. In the library accompanying this paper \cite{rotstats}, we follow \cite{bingham_glover, glover_phd} and use a lookup table and interpolate to the queried values.

{\bf Simulation:} The Bingham distribution can be simulated using an acceptance-rejection scheme (Section \ref{sec:accept_reject}) with an ACG distribution as envelope as shown in \cite{kent_mardia_simulate}. The ACG distribution is covered in Section \ref{sec:acg}. First, redefine the concentration parameters with $\kappa_i^{*}=\kappa_i-\kappa_1$ so that $\kappa_1^{*}=0$. Since $\mathcal{F}=\mathrm{B}(\mathbf{B})$ and $\mathcal{G}=\mathrm{ACG}(\mathbf{\Lambda})$, the corresponding non-normalised density functions are obtained from Eq. (\ref{eq:bingham}) and (\ref{eq:f_acg}), respectively:
\begin{equation}
\begin{split}
    & f^{*}(\mathbf{x}) = f_B^{*}(\mathbf{x}\,|\,\mathbf{B})=\exp\left(\mathbf{x}^{\top}\mathbf{B}\mathbf{x}\right)
    \\
    & g^{*}(\mathbf{x}) = f^{*}_{\mathrm{ACG}}(\mathbf{x}\,|\,\mathbf{\Lambda}) = \left(\mathbf{x}\mathbf{\Lambda}^{-1}\mathbf{x}^{\top}\right)^{-\frac{d}{2}}
\end{split}
\end{equation}

Fix the parameter of the ACG distribution to $\mathbf{\Lambda}=(\mathbf{I}_d-2\mathbf{B}^{-1}/b)^{-1}$, where $b$ satisfies:
\begin{equation}\label{eq:bingham_sam}
    \sum^d_{i=1}\frac{1}{b+2\kappa_i^{*}}=1
\end{equation}
\noindent and let $M^{*}=\exp((b-d)/2)(d/b)^{d/2}$. Using these expressions for $f^{*}$, $g^{*}$ and $M^{*}$, it is possible to use the acceptance-rejection method in Algorithm \ref{alg:acc_rej}. Note that Eq. (\ref{eq:bingham_sam}) has to be solved numerically.


\subsection{The angular central Gaussian distribution}\label{sec:acg}

\begin{figure}[ht]
    \centering
    \includegraphics[width=0.5\textwidth]{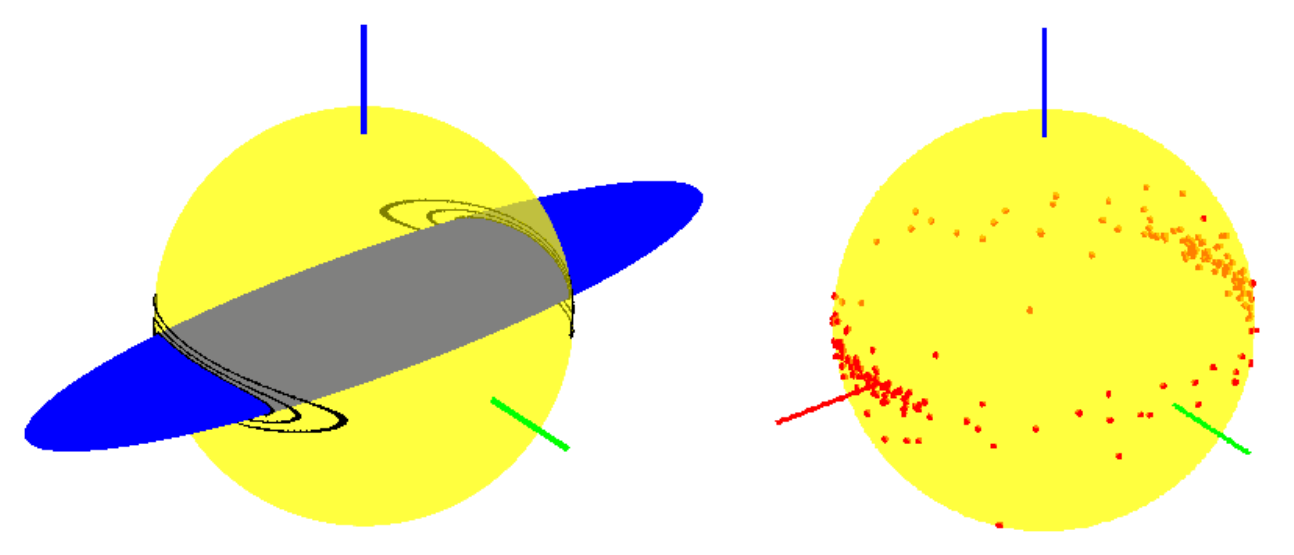}
    \caption{A $d=3$ example of an ACG distribution with $\mathbf{\Lambda}=\mathrm{diag}(4, 0.25, 0.01)$ i.e. the mean is $(1,0,0)^{\top}$ and the other principal directions are parallel to the $Y-$ and $Z-$axes. LEFT: Generation of contours by intersecting $\mathcal{S}^2$ with the contour ellipsoids of $\mathcal{N}(\mathbf{0},\mathbf{\Lambda})$, showing $(x,y,z)\mathbf{\Lambda}^{-1}(x,y,z)^{\top}=1$ in blue. RIGHT: Simulation of 200 samples.}
    \label{fig:acg}
\end{figure}

The angular central Gaussian distribution, ACG, \cite{tyler_acg} is a simpler model than the Bingham distribution but it preserves the desired properties of anti-podal symmetry and elliptical contours. The ACG distribution on $\mathcal{S}^{d-1}/\mathbb{Z}_2$ models the directions of a MVN distribution $\mathcal{N}(\mathbf{0},\mathbf{\Lambda})$ as exemplified for $d=3$ in Fig. \ref{fig:acg}. The density function is given by \cite{tyler_acg}:

\begin{equation}\label{eq:f_acg}
    f_\text{ACG}(\mathbf{x}\,|\, \mathbf{\Lambda}):= \frac{\Gamma\left(\frac{d}{2}\right)}{2\sqrt{\pi^d\left|\mathbf{\Lambda}\right|}}\left(\mathbf{x}^{\top}\mathbf{\Lambda}^{-1}\mathbf{x}\right)^{-\frac{d}{2}},\;\; \mathbf{x}\in\mathcal{S}^{d-1} /\mathbb Z_2.
\end{equation}

\noindent where $\mathbf{\Lambda}\in\mathrm{Sym}^{+}(d)$. Clearly, antipodal symmetry holds since $f_\text{ACG}(\mathbf{x}\,|\, \mathbf{\Lambda})= f_\text{ACG}(-\mathbf{x}\,|\, \mathbf{\Lambda})$. In addition, note that $f_\text{ACG}(\mathbf{x}\,|\,\mathbf{\Lambda}) = f_\text{ACG}(\mathbf{x}\,|\,c\mathbf{\Lambda})$. Hence, if $a_1\geq\ldots\geq a_d$ are the eigenvalues of $\mathbf{\Lambda}$, and $\mathbf{b}_1,\ldots,\mathbf{b}_d\in\mathcal{S}^{d-1}$ the corresponding eigenvectors, the shape of the contours only depends on the proportion of the eigenvalues $a_1,\ldots,a_d$. By convention we set $\mathrm{tr}(\mathbf{\Lambda}) = \sum_ia_i = d$. The mean is then $\mathbf{b}_1$. The ratio $a_1/a_2$ defines how concentrated the distribution is, while $a_2$ and the rest of eigenvalues define the dispersion in each principal direction.

{\bf Parameter estimation:} Condiser the dataset $\mathscr{D}=\{\mathbf{x}_i\in\mathcal{S}^{d-1}\}_{i=1}^{N}$. The MLE of the parameter $\mathbf{\Lambda}$ satisfies \cite{tyler_acg}:
\begin{equation}\label{eq:acg_mle_eq}
    \hat{\mathbf{\Lambda}} = \frac{d}{N}\sum_{i=1}^{N} \frac{\mathbf{x}_i\mathbf{x}_i^{\top}}{\mathbf{x}_i^{\top}\hat{\mathbf{\Lambda}}^{-1}\mathbf{x}_i}.
\end{equation}
In \cite{tyler_acg} it is proved that Eq. (\ref{eq:acg_mle_eq}) can be solved for $\hat{\mathbf{\Lambda}}$ by the iterative process:
\begin{equation}\label{eq:acg_mle}
    {\mathbf{\Lambda}}_{k+1} = d\displaystyle\sum_{i=1}^{N}\left\{ \dfrac{\left(\mathbf{x}_i^{\top}{\mathbf{\Lambda}}_k^{-1}\mathbf{x}_i\right)^{-1}}{\sum\limits_{i=1}^{N}\left(\mathbf{x}_i^{\top}{\mathbf{\Lambda}}_k^{-1}\mathbf{x}_i\right)^{-1}}\mathbf{x}_i\mathbf{x}_i^{\top}\right\},
\end{equation}

\noindent which begins with $\mathbf{\Lambda}_{0} = \mathbf{I}_d$. The algorithm is stopped when the difference between iterations, measured using a metric in $\mathrm{Sym}^{+}(d)$, is smaller than a fixed small value.

{\bf Simulation:} Due to the definition of the ACG distribution as the model of the directions of a zero-mean MVN distribution, to simulate $\mathrm{ACG}(\mathbf{\Lambda})$, simply draw $\mathbf{y}\sim\mathcal{N}(\mathbf{0},\mathbf{\Lambda})$, then $\mathbf{x}=\mathbf{y}/\|\mathbf{y}\|$ is ACG.


\subsection{Gaussian in the tangent space of $\mathcal{S}^{d-1}$}\label{sec:tg}

The simplest way to model orientational data is by projecting it onto the tangent space which is, by definition, a vector space. Once there, a MVN distribution can be fitted. The obvious drawback is the lack of a normalised density function and the potential distortion that the projection can bring about if the base point is not chosen correctly or if the distribution is too disperse. 

In this section we work exclusively with data in $\mathcal{S}^{d-1}$, where $d=3$ for 2-DOF axis orientation, and $d=4$ for quaternions, considering antipodal symmetry. In Sec. \ref{sec:gaussian_lie_algebra}, a similar technique is used to model data in SO(3) exploiting its Lie group structure and defining a Gaussian in its Lie algebra. While this section provides a Riemannian treatment to the modeling, Sec. \ref{sec:gaussian_lie_algebra} uses a Lie-theoretic approach.

Define a tangent space, $T_{\bm{\upmu}}\mathcal{S}^{d-1}$, at the {\it base point} $\bm{\upmu}\in\mathcal{S}^{d-1}$. Then every point on $\mathcal{S}^{d-1}$ can be projected onto the tangent space by means of the logarithmic map, $\mathrm{log}_{\bm{\upmu}}\,:\,\mathcal{S}^{d-1}\rightarrow T_{\bm{\upmu}}\mathcal{S}^{d-1}$, and points in the tangent space can be projected back to $\mathcal{S}^{d-1}$ by means of the exponential map $\mathrm{exp}_{\bm{\upmu}}\,:\,T_{\bm{\upmu}}\mathcal{S}^{d-1}\rightarrow\mathcal{S}^{d-1}$. These maps are defined as follows \cite{log_exp_sd}:
\begin{equation}\label{eq:log_exp_sd}
    \begin{split}
        \mathrm{log}_{\bm{\upmu}}(\mathbf{x})&=\frac{\theta}{\sin\theta}\left\{\mathbf{x}-\cos(\theta)\bm{\upmu}\right\}\in\mathbb{R}^d,\;\;\theta=d_{\mathcal{S}^{d-1}}(\bm{\upmu},\mathbf{x}),\;\;\mathbf{x}\in\mathcal{S}^{d-1}
        \\
        \exp_{\bm{\upmu}}(\mathbf{y})&=\cos(\|\mathbf{y}\|)\bm{\upmu} + \frac{\sin\|\mathbf{y}\|}{\|\mathbf{y}\|}\mathbf{y}\in\mathcal{S}^{d-1},\;\; \mathbf{y}\in\mathbb{R}^d
    \end{split}
\end{equation}

\noindent where $d_{\mathcal{S}^{d-1}}(\cdot,\cdot)$ is a metric in $\mathcal{S}^{d-1}$ which can be defined to force antipodal symmetry as follows \cite{calinon_gmr_manifolds}:

\begin{equation}\label{eq:d_sd}
d_{\mathcal{S}^{d-1}}(\bm{\upmu},\mathbf{x}) = 
\left\{
    \begin{array}{lr}
        \arccos(\bm{\upmu}^{\top}\mathbf{x}), & \text{if } \bm{\upmu}^{\top}\mathbf{x} \geq 0\\
        \arccos(-\bm{\upmu}^{\top}\mathbf{x}), & \text{if } \bm{\upmu}^{\top}\mathbf{x} < 0
    \end{array}
\right.
\end{equation}

If no antipodal symmetry is desired, simply define $d_{\mathcal{S}^{d-1}}(\bm{\upmu},\mathbf{x}) =\arccos(\bm{\upmu}^{\top}\mathbf{x})$. By use of Eq. (\ref{eq:d_sd}), the whole $\mathcal{S}^{d-1}$ is mapped onto a disk $\mathcal{D}_{\bm{\upmu}}$ of radius $\pi/2$. Then every point $\mathbf{y}$ in $\mathcal{D}_{\bm{\upmu}}$ maps to two points in $\mathcal{S}^{d-1}$, namely $\exp_{\bm{\upmu}}(\mathbf{y})$ and $-\exp_{\bm{\upmu}}(\mathbf{y})$.

\begin{figure}[ht]
    \centering
    \includegraphics[width=1.0\textwidth]{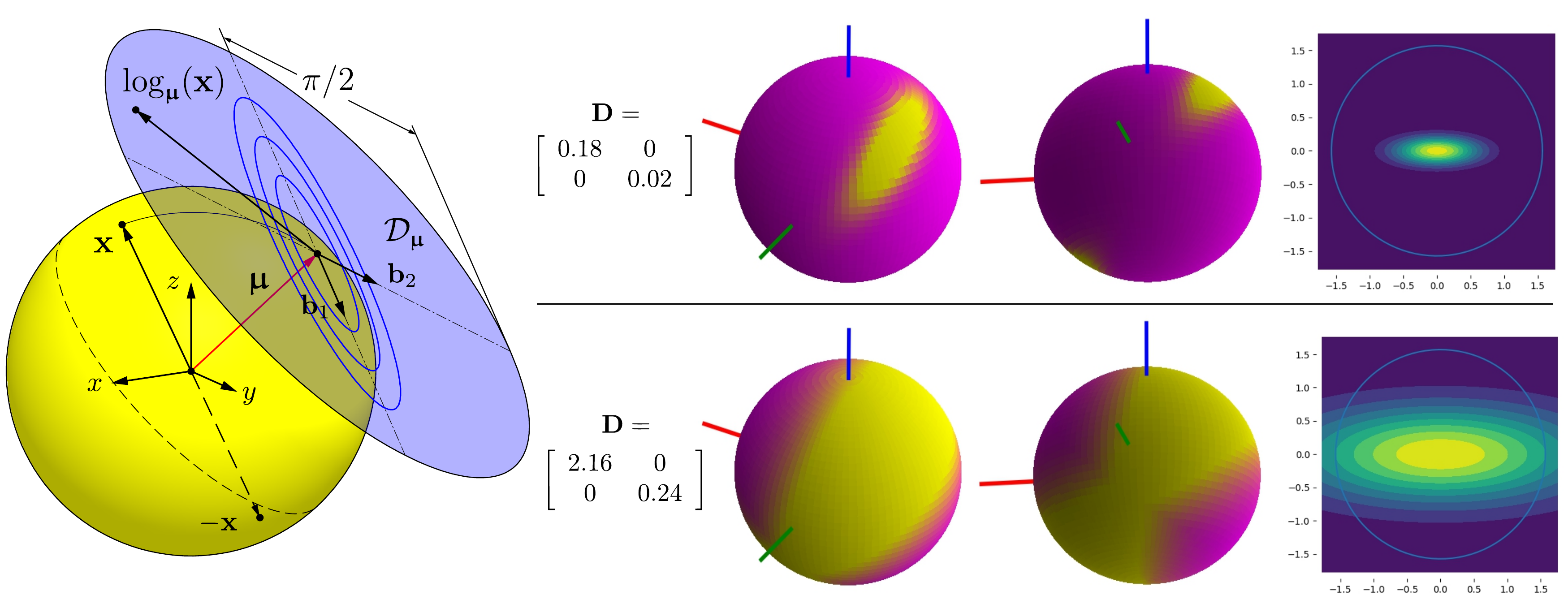}
    \caption{A MVN distribution defined in $T_{\bm{\upmu}}\mathcal{S}^2$ with $\bm{\upmu}=(-0.62,0.338, 0.707)^{\top}$, $\mathbf{b}_1=(-0.028, -0.911,0.411)^{\top}$ and $\mathbf{b}_2=\bm{\upmu}\times\mathbf{b}_1$, and with antipodal symmetry. LEFT: Schematic of the definition. RIGHT: A comparison of two distributions with different covariance, for each showing two views of $\mathcal{S}^2$ and the MVN in $T_{\bm{\upmu}}\mathcal{S}^2$. Note that, in the bottom case, the dispersion looks non-smooth across the equator.}
    \label{fig:tg}
\end{figure}

Note that the vectors in the maps in Eq. (\ref{eq:log_exp_sd}) are all in $\mathbb{R}^d$ despite the fact that $\mathrm{dim}(T_{\bm{\upmu}}\mathcal{S}^{d-1})=d-1$. Hence, before defining a MVN distribution in $T_{\bm{\upmu}}\mathcal{S}^{d-1}$ we would like to transform all the tangent vectors at $\bm{\upmu}$ to $\mathbb{R}^{d-1}$. We can pick the principal axes of the MVN as a basis for $\mathbb{R}^{d-1}$ so that the MVN distribution is $\mathcal{N}(\mathbf{0},\mathrm{diag}(\lambda_1,\ldots,\lambda_{d-1}))$ and the density function at $\mathbf{x}\in\mathcal{S}^{d-1}$ is given by (see, for example \cite{dirichlet_mixture}):
\begin{equation}\label{eq:f_TG}
    f_{\mathrm{TG},\bm{\upmu}}(\mathbf{x}\,|\,\mathbf{B},\mathbf{D}):=\frac{1}{\sqrt{(2\pi)^{d-1}\mathrm{Tr}(\mathbf{D})}}\exp\left\{-\frac{1}{2}\mathbf{\log_{\bm{\upmu}}(\mathbf{x})}^{\top}\mathbf{B}^{\top}\mathbf{D}^{-1}\mathbf{B}\log_{\bm{\upmu}}(\mathbf{x})\right\},
\end{equation}

\noindent where $\mathbf{D}=\mathrm{diag}(\lambda_1,\ldots,\lambda_{d-1})$ is the covariance matrix of the MVN, $\mathbf{B}\in\mathbb{R}^{(d-1)\times d}$ is the transformation matrix whose rows are the axes of the MVN and serve as basis for $\mathbb{R}^{d-1}$. 

Clearly, the density function (\ref{eq:f_TG}) extends outside $\mathcal{D}_{\bm{\upmu}}$ throughout $\mathbb{R}^{d-1}$ and $\int_{\mathcal{D}_{\bm{\upmu}}}f_{\mathrm{TG},\bm{\upmu}}(\mathbf{x})\mathrm{d}\mathbf{x}\neq 1$. For very concentrated data Eq. (\ref{eq:f_TG}) can give satisfactory results, hence its wide use. However, the correct normalising term can be obtained following \cite{pennec}. In \cite{spherical_normal}, a model in the tangent space with parameter $\mathbf{\Lambda}:=\mathbf{B}^{\top}\mathbf{D}^{-1}\mathbf{B}$ is introduced and expressions for the normalising constant for $d=3$ are provided. This model is known as {\it spherical normal}, SN.

\cite{spherical_normal} also suggests that distributions in the tangent space like the one in Eq. (\ref{eq:f_TG}) can be interpreted as log-normal distributions in which case the Jacobian that appears as a consequence of the change of variable is missing in most applications. The {\it log-normal distribution}, LN, is the distribution of a variable whose logarithm is normally distributed. In our case, we assume that $\mathbf{B}\mathrm{log}_{\bm{\upmu}}(\mathbf{x})\sim\mathcal{N}(\mathbf{0},\mathbf{D})$, hence $\mathbf{x}$ is LN. Due to this change of variable, the density function becomes $|\det(\mathbf{J}(\mathbf{x}))|f_{\mathrm{TG}}$, where $\mathbf{J}(\mathbf{x})=\partial_{\mathbf{x}}(\mathbf{B}\mathrm{log}_{\bm{\upmu}}(\mathbf{x}))$ is the Jacobian of the transformation. Note however that $\mathbf{J}$ depends on $\mathbf{x}$ and therefore it not only affects the normalising constant, but also the shape of the distribution.

It is important to mention a variant of Eq. (\ref{eq:f_TG}) for quaternions which is the most widely used in robot learning where it is known as {\it Riemannian Gaussian} \cite{calinon_gmr_manifolds, caldwell_projection}. In this case, Eq. (\ref{eq:so3}) becomes:
\begin{equation}\label{eq:f_RG}
    f_{\mathrm{RG}}(\mathbf{q}\,|\,\bm{\upmu},\mathbf{\Sigma}):=\frac{1}{\sqrt{(2\pi)^3\det(\mathbf{\Sigma})}}\exp\left[-\frac{1}{2}\left(\log(\bm{\upmu}^{-1}*\mathbf{q})\right)^{\top}\mathbf{\Sigma}^{-1}\log(\bm{\upmu}^{-1}*\mathbf{q})\right],
\end{equation}
\noindent where $\mathbf{q},\bm{\upmu}\in\mathcal{S}^{3}/\mathbb{Z}_2$, $\mathbf{\Sigma}\in\mathrm{Sym}^{+}(3)$, and the log map is given by $\log(\mathbf{q})=d_{\mathcal{S}^3}\mathbf{v}/\|\mathbf{v}\|$, where, if $\mathbf{q}=(x,y,z,w)^{\top}$, then $\mathbf{v}=(x,y,z)^{\top}$. Note that this definition of log is exclusive for quaternions, using their kinematic meaning \--- the result, sometimes called the {\it rotation vector} \cite{shuster_survey_repr}, is in $\mathbb{R}^3$ and is the quaternion equivalent of the log in Eq. (\ref{eq:exp_log_so3}) defined in the Lie algebra of SO(3). In contrast, the log in Eq. (\ref{eq:log_exp_sd}) is based on the geometry of $\mathcal{S}^{d-1}$ for any $d$, and, if applied to a quaternion ($d=4$), the result is in $\mathbb{R}^4$. Hence, Eq. (\ref{eq:f_RG}) is simply the quaternion version of Eq. (\ref{eq:so3}), which is a Gaussian defined in the Lie algebra of SO(3).

{\bf Parameter estimation:} Key to minimising distortion when projecting the data onto the tangent space is that the base point $\bm{\upmu}$ be the mean. In this case, we are free to choose any definition of the mean since the model does not provide information about it. When working with a dataset $\mathscr{D}=\{\mathbf{x}_i\in\mathcal{S}^{d-1}/\mathbb{Z}_2\}_{i=1}^N$ where antipodal symmetry is present, we can proceed as we did in Section (\ref{sec:bingham}) and estimate the mean as the eigenvector corresponding to the largest eigenvalue of the scatter matrix $\mathbf{S}$. 

Alternatively, the base point can also be estimated in a Riemannian way by means of the {\it Fr\`echet mean} (see, for example \cite{geo_mean_fletcher}):
\begin{equation}\label{eq:frechet}
    \hat{\bm{\upmu}} = \arg \min_{\bm{\upmu}\in\mathcal{S}^{d-1}}\frac{1}{N}\sum_{i=1}^Nd_{\mathcal{S}^{d-1}}(\bm{\upmu},\mathbf{x}_i)
\end{equation}
\noindent which can be found with the iterative scheme starting with $\bm{\upmu}_0=\mathbf{x}_1$ \cite{geo_mean_fletcher}:
\begin{equation}\label{eq:mu_k}
        \bm{\upmu}_{k+1}=\exp_{\bm{\upmu}_{k}}\left[\frac{1}{N}\sum_{i=1}^N\log_{\bm{\upmu}_{k}}\left(\mathbf{x}_i\right)\right]
\end{equation}
Note that, in general, these two methods of computing the mean do not generate equivalent results.

To find $\hat{\mathbf{B}}$ consider: 
\[
    \mathbf{L}:=\left[\log_{\hat{\bm{\upmu}}}(\mathbf{x}_i)^{\top},\ldots,\log_{\hat{\bm{\upmu}}}(\mathbf{x}_N)^{\top}\right]^{\top}\in\mathbb{R}^{N\times d}
\]
Then the principal directions of the tangent vectors can be found by singular value decomposition (SVD) $\mathbf{L}=\mathbf{U}_{L}\mathbf{D}_{L}\mathbf{V}_{L}^{\top}$, where it is assumed that the singular values appear in descending order in $\mathbf{D}_{L}$. Then, if $\mathbf{V}_{L}=[\mathbf{e}_1,\ldots,\mathbf{e}_d]$, it follows that $\hat{\mathbf{B}}=[\mathbf{e}_1,\ldots,\mathbf{e}_{d-1}]^{\top}\in\mathbb{R}^{(d-1)\times d}$.

Finally, a $(d-1)-$dimensional, zero-mean MVN distribution with diagonal covariance matrix is fitted to the dataset $\{\hat{\mathbf{B}}\log_{\hat{\bm{\upmu}}}(\mathbf{x}_i)\in\mathbb{R}^{d-1}\}^N_{i=0}$. Thus estimating $\hat{\lambda}_1\ldots\hat{\lambda}_{d-1}$.

{\bf Simulation:} If the distribution is concentrated, it is possible to sample from the MVN and use the exponential map to return the sample to $\mathcal{S}^{d-1}$ after considering rejection outside $\mathcal{D}_{\bm{\upmu}}$.


\section{3-DOF orientations: Rotation matrices, SO(3)}\label{sec:SO3}

\noindent This section includes models for rotation matrices. The model presented in \ref{sec:gaussian_lie_algebra} is commonly used for quaternions too. Hence, a quaternion version of the equations is also presented.

\subsection{The matrix Fisher distribution}\label{sec:MF}

The matrix Fisher distribution \cite{MF1,MF2}, MF, is a distribution on SO(3) and is closely related to the Bingham distribution \cite{prentice_bingham_matfisher}. In this section we follow the notation in \cite{lee_matF}. This model is a special case of a more general family of distributions on Stiefel manifolds \cite{kume_saddle_stiefel}.The density function of the MF distribution at $\mathbf{R}\in\mathrm{SO}(3)$ is given by \cite{MF1,MF2}:
\begin{equation}\label{eq:f_MF}
    f_{\mathrm{MF}}(\mathbf{R}\,|\,\mathbf{F}):=\frac{1}{c(\mathbf{F})}\exp\left\{\mathrm{Tr}\left(\mathbf{F}^{\top}\mathbf{R}\right)\right\}
\end{equation}

\noindent where $\mathbf{F}\in\mathbb{R}^{3\times 3}$ has SVD $\mathbf{F}=\mathbf{U}'\mathbf{\Psi}'(\mathbf{V}')^{\top}$, with singular values $\psi_1'\leq \psi_2'\leq \psi_3'$. To ensure that matrices of singular vectors are in SO(3), the {\it proper} SVD is defined as $\mathbf{F}=\mathbf{U}\mathbf{\Psi}\mathbf{V}^{\top}$, where:
\begin{equation}
    \begin{split}
        \mathbf{U} &= \mathbf{U}'\mathrm{diag}(1,1,\det(\mathbf{U}'))\in\mathrm{SO}(3)\\
        \mathbf{\Psi} &= \mathrm{diag}(\psi_1,\psi_2,\psi_3)=\mathrm{diag}(\psi_1',\psi_2',\det(\mathbf{U}'\mathbf{V}')\psi_3')\\
        \mathbf{V} &= \mathbf{V}'\mathrm{diag}(1,1,\det(\mathbf{V}'))\in\mathrm{SO}(3)
    \end{split}
\end{equation}

The normalising term, $c(\mathbf{F})$, depends only on $\mathbf{\Psi}$, and is given by \cite{lee_matF}:
\begin{equation}\label{eq:MF_c}
    c(\mathbf{\Psi})=\frac{1}{2}\int^1_{-1}I_0\big[A_1(\mathbf{\Psi},u)\big]I_0\big[A_2(\mathbf{\Psi},u)\big]\exp(\psi_ku)\mathrm{d}u
\end{equation}

\noindent where $(i,j,k)\in\{(1,2,3),(2,3,1),(3,1,2)\}$, $A_1(\mathbf{\Psi},u)=(\psi_i-\psi_j)(1-u)/2$, $A_2(\mathbf{\Psi},u)=(\psi_i+\psi_j)(1+u)/2$, and $I_0$ and $I_1$ are the modified Bessel functions of the first kind and orders 0 and 1, respectively (see Eq. (\ref{eq:I0_I1})).

\begin{figure}[ht]
    \centering
    \includegraphics[width=0.55\textwidth]{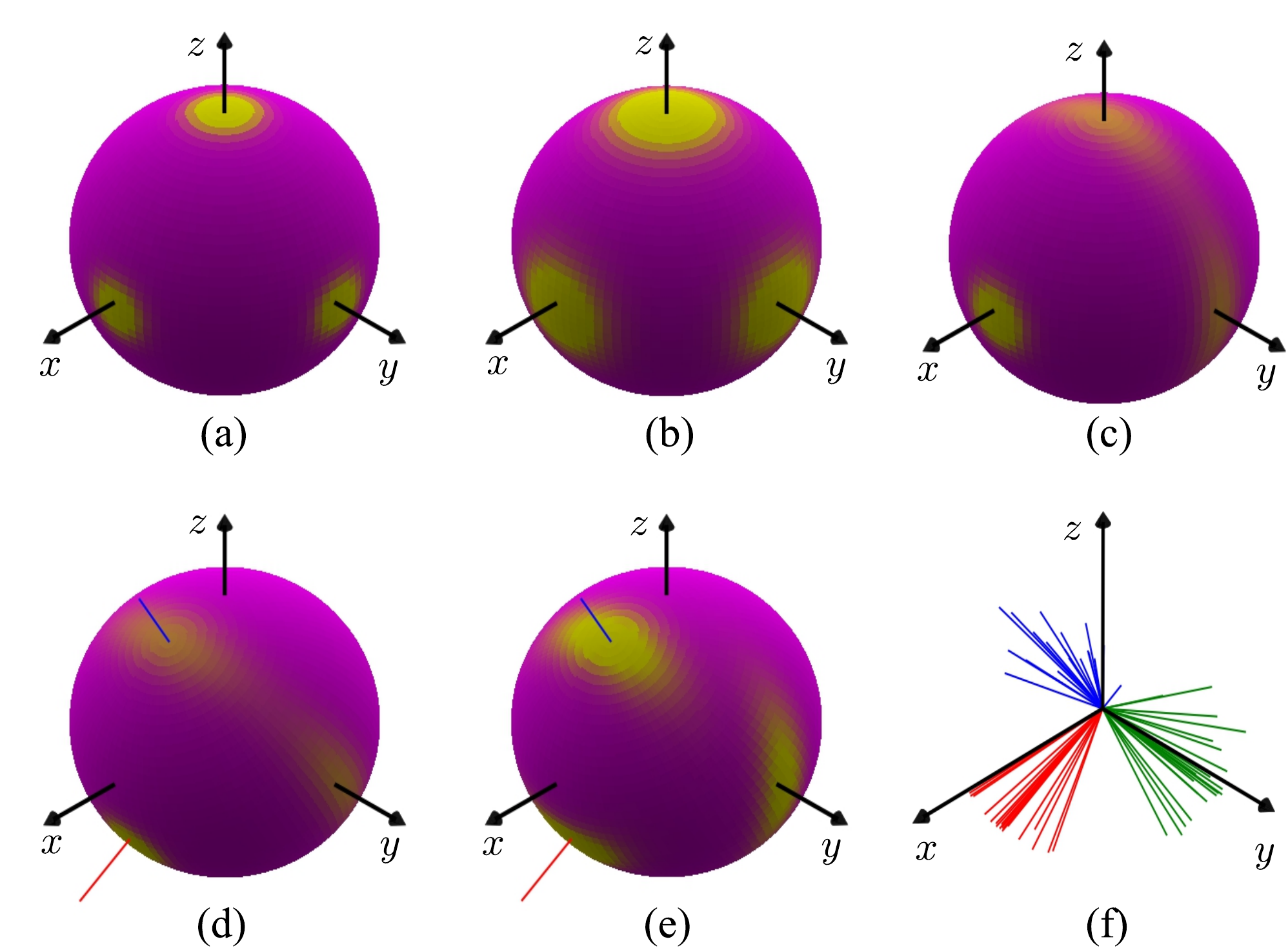}
    \caption{Visualisation of MF($\mathbf{F}$) a) $\mathbf{F}=10\mathbf{I}_3$, b) $\mathbf{F}=30\mathbf{I}_3$, c) $\mathbf{F}=\mathrm{diag}(45,5,1)$, d) $\mathbf{F}=\exp((0,\pi/6,0)^{\top})\mathrm{diag}(45,5,1)$, e) $\mathbf{F}=\mathrm{diag}(45,5,1)\exp((0,\pi/6,0)^{\top})$, f) 20 samples from the distribution in e)}
    \label{fig:mf}
\end{figure}

In this model, the mean is given by $\mathbf{M}=\mathbf{U}\mathbf{V}^{\top}\in\mathrm{SO}(3)$. The columns of $\mathbf{U}$ are the principal axes of rotation and $\{\psi_i,\psi_j,\psi_k\}$ give the concentration in the directions of $\mathbf{U}$. The larger the concentration around the axis $\mathbf{M}\mathbf{e}_i$ the larger the quantity $\psi_j+\psi_k$. Fig. \ref{fig:mf} shows the influence of this parameters in the shape of the distribution.

{\bf Parameter estimation:} Given the dataset $\mathscr{D}=\{\mathbf{R}_i\in\mathrm{SO}(3)\}_{i=1}^N$, consider the SVD of the sample mean:
\begin{equation}\label{eq:R_samplemean}
    \overline{\mathbf{R}}=\frac{1}{N}\sum^N_{i=1}\mathbf{R}_i=\mathbf{U}_R\mathbf{\Psi}_R(\mathbf{V}_R)^{\top}
\end{equation}
\noindent then $\hat{\mathbf{U}}=\mathbf{U}_R$ and $\hat{\mathbf{V}}=\mathbf{V}_R$ \cite{lee_matF}. 

Let $\mathbf{\Psi}_R=\mathrm{diag}(d_1,d_2,d_3)$, then the MLE of $\mathbf{\Psi}=\mathrm{diag}(\psi_1,\psi_2,\psi_3)$ satisfies:
\begin{equation}\label{eq:MF_mle}
    \frac{1}{c(\hat{\mathbf{\Psi}})}\left[\frac{\partial c(\hat{\mathbf{\Psi}})}{\partial \psi_i}\right] = d_i,\:\; i=1,2,3.
\end{equation}

\noindent where the partial derivatives of the normalising constant are given by \cite{lee_matF}:
\begin{equation}\label{eq:MF_dc}
\begin{split}
    \frac{\partial c(\mathbf{\Psi})}{\partial \psi_i} =& \frac{1}{4}\int^1_{-1}\bigg\{(1-u)I_1\big[A_1(\mathbf{\Psi},u)\big] I_0\big[A_2(\mathbf{\Psi},u)\big]+(1+u)I_0\big[A_1(\mathbf{\Psi},u)\big]I_1\big[A_2(\mathbf{\Psi},u)\big]\bigg\}\exp(\psi_ku)\mathrm{d}u,
    \\
    \frac{\partial c(\mathbf{\Psi})}{\partial \psi_j} =& \frac{1}{4}\int^1_{-1}\bigg\{-(1-u)I_1\big[A_1(\mathbf{\Psi},u)\big] I_0\big[A_2(\mathbf{\Psi},u)\big]+(1+u)I_0\big[A_1(\mathbf{\Psi},u)\big]I_1\big[A_2(\mathbf{\Psi},u)\big]\bigg\}\exp(\psi_ku)\mathrm{d}u,
    \\
    \frac{\partial c(\mathbf{\Psi})}{\partial \psi_k} =& \frac{1}{2}\int^1_{-1}(1-u)I_0\big[A_1(\mathbf{\Psi},u)\big] I_0\big[A_2(\mathbf{\Psi},u)\big]u\exp(\psi_ku)\mathrm{d}u
\end{split}
\end{equation}

Eq. (\ref{eq:MF_mle}) is a system of three equations in the unknowns $\{\hat{\psi}_1,\hat{\psi}_2,\hat{\psi}_3\}$, which are the remaining estimates to fit a MF distribution to $\mathscr{D}$. This system of equations has to be solved numerically. The computationally expensive part of this task is the evaluation of the integrals in the normalising constant (Eq. (\ref{eq:MF_c})) and its derivatives (Eq. (\ref{eq:MF_dc})), which have to be computed numerically. In the Python library accompanying this paper \cite{rotstats}, each integrand is defined as a \texttt{Cython} function. \texttt{Cython} fully converts the integrand into a \texttt{C} function that is callable from \texttt{SciPy}. This increases the performance by around 10 times.

{\bf Simulation:} An easy way to simulate the MF distribution is by sampling from the Bingham distribution and then using their relationship to transform the samples into MF samples \cite{lee_matF}. Given MF($\mathbf{F}$), with proper SVD $\mathbf{F}=\mathbf{U}\mathbf{\Psi}\mathbf{V}^{\top}$, set:
\begin{equation}
    \mathbf{B} = \mathrm{diag}\big\{2\psi_1-\mathrm{Tr}(\mathbf{\Psi}),\, 2\psi_2-\mathrm{Tr}(\mathbf{\Psi}),\, 2\psi_3-\mathrm{Tr}(\mathbf{\Psi}),\, \mathrm{Tr}(\mathbf{\Psi})\big\}
\end{equation}

Then, using the method described in Section \ref{sec:bingham}, simulate $\mathbf{q}\sim\mathrm{B}(\mathbf{B})$ where $\mathbf{q}\in\mathcal{S}^3/\mathbb{Z}_2$. The samples $\mathbf{q}$ are quaternions that can be converted into rotation matrices by means Eq. (\ref{eq:rodrigues}), $\mathbf{R}_B=\mathrm{Rod}(\mathbf{q})\in\mathrm{SO}(3)$. Finally, reorientate the sample to the original coordinate system: $\mathbf{R}=\mathbf{U}\mathbf{R}_B\mathbf{V}^{\top}$, then $\mathbf{R}\sim\mathrm{MF}(\mathbf{F})$. Note that with this technique, the underlying simulated distribution is actually an ACG since we simulate the Bingham distribution through acceptance-rejection with an ACG envelope.

\subsection{Gaussian in the Lie algebra of SO(3)}\label{sec:gaussian_lie_algebra}

In Sec. \ref{sec:tg}, a Gaussian was defined in a tangent space of $\mathcal{S}^{d-1}$. In this section, the same approach is used to define a Gaussian in the Lie algebra of SO(3) when seen as a Lie group.

Let $G$ be a Lie group with operation $\circ$, then it is possible to establish a Gaussian in its Lie algebra $\mathfrak{g}$ which is, by definition, a vector space. Since $\mathfrak{g}$ is only defined at the identity, the data must be left-multiplied by the inverse element of the base point $\mu\in G$, resulting in the following density function at $g\in G$ \cite{greg_nonparametric}:
\begin{equation}\label{eq:fg}
    f_{\mathfrak{g}}(g\,|\,\mu,\mathbf{\Sigma}):=\frac{1}{c(\mathbf{\Sigma})}\exp\left[-\frac{1}{2}\left(\log(\mu^{-1}\circ g)^{\vee}\right)^{\top}\mathbf{\Sigma}^{-1}\log(\mu^{-1}\circ g)^{\vee}\right]
\end{equation}

\noindent where $(\cdot)^{\vee}$ is the vector version of elements in $\mathfrak{g}$. For the case of SE(2), this model has been nicknamed {\it the banana distribution} \cite{banana} due to the shape of the distribution of the tip of the heading of a differential robot moving on the plane. In the case of SO(3), the data must be re-oriented with respect to the mean orientation and (\ref{eq:fg}) becomes for $\mathbf{R}\in\mathrm{SO}(3)$:
\begin{equation}\label{eq:so3}
    f_{\mathfrak{so}(3)}(\mathbf{R}\,|\,\mathbf{M},\mathbf{\Sigma}):=\frac{1}{c(\mathbf{\Sigma})}\exp\left[-\frac{1}{2}\left(\log(\mathbf{M}^{-1}\mathbf{R})^{\vee}\right)^{\top}\mathbf{\Sigma}^{-1}\log(\mathbf{M}^{-1}\mathbf{R}))^{\vee}\right],
\end{equation}

\noindent where $\mathbf{M}\in\mathrm{SO}(3)$ is the mean and $\mathbf{\Sigma}\in\mathrm{Sym}^{+}(3)$ is the covariance. The corresponding log and exp maps are defined as follows (see Sec. 4.4 of \cite{selig}):
\begin{equation}\label{eq:exp_log_so3}
    \begin{split}
        \exp(\bm{\upomega})&=\mathbf{I}_3+\frac{\sin\|\bm{\upomega}\|}{\|\bm{\upomega}\|}\mathrm{skew}(\bm{\upomega})+\frac{1-\cos\|\bm{\upomega}\|}{\|\bm{\upomega}\|^2}\mathrm{skew}(\bm{\upomega})^2 \in \mathrm{SO}(3),
        \\
        \log(\mathbf{R})&=
        \left\{\begin{array}{ll}
        \dfrac{\theta}{2\sin\theta}\left(\mathbf{R}-\mathbf{R}^{\top}\right) \in \mathfrak{so}(3), & \text{if } \theta\notin \{0,\pi\}\\
        \mathbf{0}_{3\times 3}, & \text{if } \theta=0
        \end{array}
        \right.\\
        \text{with  }\theta &= \arccos((\mathrm{Tr}(\mathbf{R})-1)/2)
    \end{split}
\end{equation}
\noindent where $\mathbf{R}\in\mathrm{SO}(3)$ and $\mathrm{skew}(\bm{\upomega})\in\mathfrak{so}(3)$ with $\mathrm{skew}(\cdot)$ defined in Eq. (\ref{eq:skew}). The vector $\bm{\upomega}=(\log(\mathbf{R}))^{\vee}\in\mathbb{R}^3$ is the angular velocity needed to rotate an object to an orientation given by $\mathbf{R}$ in a unit of time. Such rotation has axis $\bm{\upomega}/\|\bm{\upomega}\|$ and speed magnitude $\|\bm{\upomega}\|$. The latter can also be seen as the total angle of displacement and $\mathbf{r}=\bm{\upomega}$ is then called {\it rotation vector}. Note that the exponential map in (\ref{eq:exp_log_so3}) is a consequence of the Rodrigues' formula in Eq. (\ref{eq:rodrigues}).

For very concentrated data, the normalising constant is approximated to that of a MVN, and $c(\mathbf{\Sigma})=(2\pi)^{3/2}|\det\mathbf{\Sigma}|^{1/2}$.

{\bf Parameter estimation:} Similarly to Section \ref{sec:tg}, in this model we are free to choose any definition for the mean which will serve as base point. For a dataset $\mathscr{D}=\{\mathbf{R}_i\in\mathrm{SO}(3)\}_{i=1}^N$, an easy option is to proceed as in Section \ref{sec:MF} and estimate the mean using the SVD of the sample mean in Eq. (\ref{eq:R_samplemean}), $\overline{\mathbf{R}}=\mathbf{U}_R\mathbf{\Psi}\mathbf{V}_R^{\top}$, then $\hat{\mathbf{M}}=\mathbf{U}_R\mathbf{V}_R^{\top}$. Such a mean is the closest rotation matrix to the sample mean and coincides with the MLE of the MF distribution.

Alternatively, for a general group $G$, $\sum_i\log(\hat{\mu}^{-1}\circ g_i)=\mathbf{0}$, $g_i\in G$, which ensures that the Gaussian fitted in the Lie algebra has zero mean, can be solved with a Lie-theoretic equivalent of Eq. (\ref{eq:mu_k}), as shown in \cite{banana}. In the case of SO(3), it follows that:
\[
    \mathbf{M}_{k+1} = \mathbf{M}_k\mathrm{exp}\left[\frac{1}{N}\sum_{i=1}^N\log\left(\mathbf{M}_k^{\top}\mathbf{R}_i\right)\right]
\]

These two means are different in the general case and, in the Riemannian case, the result also depends on the choice of metric.

The covariance $\mathbf{\Sigma}$ is obtained as usually done for a MVN, after data projection:
\[
    \hat{\mathbf{\Sigma}}=\frac{1}{N}\sum^{N}_{i=1}\log(\hat{\mathbf{M}}^{-1}\mathbf{R}_i)^{\vee}\left(\log(\hat{\mathbf{M}}^{-1}\mathbf{R}_i)^{\vee}\right)^{\top}
\]

{\bf Simulation:} Similarly to the Gaussian defined in the tangent space of $\mathcal{S}^{d-1}$ in Section \ref{sec:tg}, sampling from the MVN defined in the Lie algebra and then using the exponential map to send the samples to SO(3) is only recommended when the concentration is high and rejection outside the ball of radius $\pi/2$ is considered.


\section{Probability density functions as solutions to diffusion equations}\label{sec:diffusion}

The models presented in Sections \ref{sec:1dof} to \ref{sec:SO3} are designed by establishing parameters that describe geometric properties of the space where the data live. A different way to generate PDFs is by solving diffusion equations. These are partial differential equations that govern how a property, like heat or density, spread with time across a medium. We can consider such property to be the probability density, and the medium, the data space. In this section, expressions for diffusion equations in $\mathbb{R}^n$, $\mathbb{S}$, $\mathcal{S}^2$ and SO(3), as well as their solutions, are reviewed. 

Note that this section intends to be a brief survey and thus fitting and sampling are not addressed. Similarly, several aspects, like the fascinating properties of the spherical harmonics and the unitary irreducible representations of SO(3), are not discussed. For a detailed treatment, the reader is addressed to \cite{greg_noncommutative_book, greg_stochastic_2}. Due to the illustrative character of this section, only solutions for isotropic diffusion are presented. However, those solutions are paramount in wavelet analysis and non-parametric density estimation \cite{laplacian_SO3, diffusion_S2_3, diffusion_S2_4, density_estimation_kernel_SO3}, where they are used as kernels. More complex diffusion models can be found in the literature on liquid crystals \cite{crystal_liq_deSouza, crystal_liq_tarroni} and polymers \cite{polymers_zero}.

Consider the general diffusion equation in Euclidean space for $f(\mathbf{x},t)$, $\mathbf{x}:=(x_1,\ldots, x_n)^{\top}\in\mathbb{R}^n$ \cite{greg_stochastic_1}:
\begin{equation}\label{eq:diffusion_euclidean}
    \frac{\partial f(\mathbf{x},t)}{\partial t}=-\triangledown\cdot\left(\mathbf{A}f(\mathbf{x},t)\right)+\triangledown\cdot\left(\mathbf{D}\triangledown f(\mathbf{x},t)\right)=-\sum^n_{i=1}A_i\frac{\partial f(\mathbf{x},t)}{\partial x_i}+\sum^n_{i=1}\sum^n_{j=1}D_{ij}\frac{\partial f^2(\mathbf{x},t)}{\partial x_i\partial x_j}
\end{equation}

\noindent where $D_{ij}$ and $A_i$ are the entries of the {\it diffusion tensor}, $\mathbf{D}\in\mathrm{Sym}(n)$, and the {\it drift vector}, $\mathbf{A}\in\mathbb{R}^n$, respectively. If $\mathbf{A}=\mathbf{0}$, $\bm{\Sigma}=2t\mathbf{D}\in\mathrm{Sym}^{+}(n)$, and the initial solution $f(\mathbf{x},0)=\delta(\mathbf{x}-\bm{\upmu})$ is used, then the solution to the diffusion equation (\ref{eq:diffusion_euclidean}) is the expression for the PDF of the MVN, $\mathcal{N}(\bm{\upmu},\bm{\Sigma})$. Since the diffusion equation preserves the mass, and $\int \delta(\mathbf{x}-\bm{\upmu})\mathrm{d}\mathbf{x}=1$, it is ensured that $f$ is a PDF at any $t\geq 0$. The addition of a drift term when $\mathbf{A}\neq\mathbf{0}$ generalises the MVN adding bias and deforming the elliptical contours in the direction of $\mathbf{A}$.

In the previous example, the medium of diffusion is $\mathbb{R}^n$. Now consider the case in which the medium is the circle, $\mathbb{S}$. Consider the following diffusion equation without drift for $f(\theta,t)$, $\theta\in\mathbb{S}$:

\begin{equation}\label{eq:diffusion_circle}
    \frac{\partial f(\theta,t)}{\partial t} = D\triangledown^2 f(\theta,t)=D\frac{\partial^2f(\theta,t)}{\partial\theta^2}
\end{equation}

Writing the solution as a Fourier series and taking the initial condition $f(\theta,0)=\delta(\theta-\mu)$, the following solution to (\ref{eq:diffusion_circle}) is found \cite{greg_stochastic_1, greg_noncommutative_book}:

\begin{equation}
    f(\theta,t) = \frac{1}{2\pi}+\frac{1}{\pi}\sum^{\infty}_{n=1}\mathrm{e}^{-Dtn^2}\cos(n(\theta-\mu))
\end{equation}

This solution is equivalent to the PDF of the wrapped normal distribution in Eq. (\ref{eq:f_wn}) with $\sigma^2=2Dt$.

Now consider the diffusion equation on the unit sphere for $f(\mathbf{u},t)$, where $\mathbf{u}$ $=$ $(\cos\varphi\sin\vartheta,$ $\sin\varphi\sin\vartheta,\cos\vartheta )^{\top}\in\mathcal{S}^2$:
\begin{equation}\label{eq:diffusion_s2}
    \frac{\partial f(\mathbf{u},t)}{\partial t} = D_{\vartheta}\frac{1}{\sin \vartheta} \frac{\partial}{\partial \vartheta} \left( \sin \vartheta \frac{\partial f(\mathbf{u},t)}{\partial \vartheta} \right) + D_{\varphi}\frac{1}{\sin^2 \vartheta} \frac{\partial^2f(\mathbf{u},t)}{\partial \varphi^2}
\end{equation}

\noindent where $D_{\vartheta}$ and $D_{\varphi}$ are the diffusion coefficients in the directions of $\vartheta$ and $\varphi$, respectively. Although (\ref{eq:diffusion_s2}) provides an anisotropic diffusion, the principal axes of the contours are fixed. However, the introduction of mixed partial derivatives can be avoided by manually rotating the model. A more common diffusion equation considers fully isotropic diffusion with $D_{\vartheta}=D_{\varphi}=D$. Then, Eq.(\ref{eq:diffusion_s2}) is usually written as $\partial f/\partial t = D\Delta_{\mathcal{S}^2}f(\mathbf{u},t)$, where $\Delta_{\mathcal{S}^2}$ is the Laplace-Beltrami operator on $\mathcal{S}^2$. To solve this diffusion equation, write the solution in the form of a Fourier series in $\mathcal{S}^2$ \cite{bulow_diffusion_S2,greg_noncommutative_book}:
\[
f(\mathbf{u},t)=\sum_{l=0}^{\infty}\sum_{m=-l}^lf_{lm}Y^m_l(\mathbf{u})
\]
\noindent where $f_{lm}$ are the Fourier coefficients and $Y^m_l:\mathcal{S}^2\rightarrow\mathbb{C}$ are the {\it spherical harmonics} given by:
\[
Y^m_l(\mathbf{u}(\vartheta,\varphi)):=\sqrt{\frac{2l+1}{4\pi}\frac{(l-m)!}{(l+m)!}}P^m_l(\cos\vartheta)\mathrm{e}^{im\varphi}
\]
\noindent where $P^m_l$ are the {\it Legendre functions}:
\[
P^m_l(x)=\frac{(-1)^m(1-x^2)^{m/2}}{2^ll!}\frac{\mathrm{d}^{l+m}}{\mathrm{d}x^{l+m}}(x^2-1)^l
\]
It can be shown \cite{bulow_diffusion_S2,diffusion_S2_2,diffusion_S2_3,diffusion_S2_4} that the initial condition $f(\mathbf{u},0)=\delta_{\mathcal{S}^2}(\mathbf{u},\bm{\upmu})$, where the Dirac delta in the sphere is defined as $\delta_{\mathcal{S}^2}(\mathbf{u}(\vartheta,\varphi),\bm{\upmu}(\vartheta_{\bm{\upmu}},\varphi_{\bm{\upmu}})):=\delta(\cos\vartheta-\cos\vartheta_{\bm{\upmu}})\delta(\varphi-\varphi_{\bm{\upmu}})$, leads to the following solution for the fully isotropic case:
\[
f(\mathbf{u},t)=\sum_{l=0}^{\infty}\sum_{m=-l}^l\mathrm{e}^{-l(l+1)Dt}Y^m_l(\mathbf{u})Y^m_l(\bm{\upmu})
\]
Now consider the general diffusion equation in SO(3) for $f(\mathbf{R},t)$, $\mathbf{R}\in\mathrm{SO}(3)$ \cite{greg_stochastic_2}:
\begin{equation}\label{eq:diffusion_so3}
    \frac{\partial f(\mathbf{R},t)}{\partial t}=-\sum^3_{i=1}A_i\tilde{X}_if(\mathbf{R},t)+\sum^3_{i=1}\sum^3_{j=1}D_{ij}\tilde{X}_i\tilde{X}_jf(\mathbf{R},t)
\end{equation}
\noindent where the differential operators $\tilde{X}_i$ for SO(3) are given by \cite{greg_stochastic_2}:
\[
\tilde{X}_1=\frac{\sin\gamma}{\sin\beta}\frac{\partial}{\partial\alpha}+\cos\gamma\frac{\partial}{\partial\beta}-\cot\beta\sin\gamma\frac{\partial}{\partial\gamma},\;\;\;
\tilde{X}_2=\frac{\cos\gamma}{\sin\beta}\frac{\partial}{\partial\alpha}-\sin\gamma\frac{\partial}{\partial\beta}-\cot\beta\cos\gamma\frac{\partial}{\partial\gamma},\;\;\;\tilde{X}_3=\frac{\partial}{\partial\gamma}
\]
where $\{\alpha,\beta,\gamma\}$ are the ZXZ Euler angles. Since SO(3) is not an abelian group, solutions to differential equations of this kind are studied by the field of {\it noncommutative harmonic analysis}. Solutions are written as a Fourier series in the group, in this case SO(3) \cite{greg_stochastic_2,density_estimation_kernel_SO3}:
\[
f(\mathbf{R},t)=\sum_{l=0}^{\infty}\sum_{m=-l}^l\sum_{n=-l}^l(2l+1)f^l_{mn}D^l_{mn}(\mathbf{R})
\]
\noindent where $f^l_{mn}$ are the Fourier coefficients, and $D^l_{mn}(\mathbf{R}):\mathrm{SO}(3)\rightarrow\mathbb{C}$ are the matrix elements of the {\it irreducible unitary representation} (IUR) of SO(3), also known as the {\it Wigner D-functions}, which are given by:
\[
    D^l_{mn}(\mathbf{R}(\alpha,\beta,\gamma))=\mathrm{e}^{-im\alpha}P^l_{mn}(\cos\beta)\mathrm{e}^{-in\gamma}
\]
If the diffusion is fully isotropic with no drift, i.e. $A_i=0$, $D_{ij}=0$ for $i\neq j$, and $D_{ii}=D$ in Eq. (\ref{eq:diffusion_so3}), then the diffusion equation is written as $\partial f(\mathbf{R},t)/\partial t = D\Delta_{\mathrm{SO}(3)}f(\mathbf{R},t)$, where $\Delta_{\mathrm{SO}(3)}=\tilde{X}_1^2+\tilde{X}_2^2+\tilde{X}_3^2$ is the Laplace-Beltrami operator on SO(3). The solution for this case is obtained in, for example, \cite{laplacian_SO3,density_estimation_kernel_SO3,gregs_student_wang}.  The initial condition $f(\mathbf{R},0)=\delta_{\mathrm{SO}(3)}(\mathbf{M}^{\top}\mathbf{R})$ is used, where $\delta_{\mathrm{SO(3)}}(\cdot)$ is the Dirac function in SO(3), see Sec. 9.2.1 of \cite{greg_noncommutative_book}. Due to the isotropy of this case, the solution can be expressed in terms of the radially-symmetric {\it zonal functions}, which are defined as:
\[
\mathcal{X}^l(\mathbf{R}):=\sum^l_{n=-l}D^l_{nn}(\mathbf{R})=\frac{\sin\left(\left(l+\frac{1}{2}\right)\theta(\mathbf{R})\right)}{\sin\left(\frac{1}{2}\theta(\mathbf{R})\right)},\;\;\theta(\mathbf{R})=\arccos\left(\frac{1}{2}\mathrm{Tr}((\mathbf{R})-1)\right)
\]
\noindent It follows that \cite{laplacian_SO3}:
\begin{equation}\label{eq:diffusion_so3_sol}
    f(\mathbf{R},t)=1+\sum_{l=1}^{\infty}(2l+1)\mathrm{e}^{-l(l+1)Dt}\mathcal{X}_l(\mathbf{R})\mathcal{X}_l(\mathbf{M})
\end{equation}
Note that, Eq. (\ref{eq:diffusion_so3_sol}) is isotropic around $\mathbf{M}$ since only the rotation angle $\theta$ appears in this solution. Hence, $f(\mathbf{R}_1,t)=f(\mathbf{R}_2,t)$ if $\theta(\mathbf{M}^{\top}\mathbf{R}_1)=\theta(\mathbf{M}^{\top}\mathbf{R}_2)$.


\section{Examples with real data}\label{sec:experiments}

\noindent In this section, some of the models discussed in this paper are applied to two different datasets. These results are obtained using the Python library accompanying the paper \cite{rotstats}.

\subsection{Experiment 1: Demonstrations of a pouring task}\label{sec:exp_listerine}

\begin{figure}[ht]
    \centering
    \includegraphics[width=0.95\textwidth]{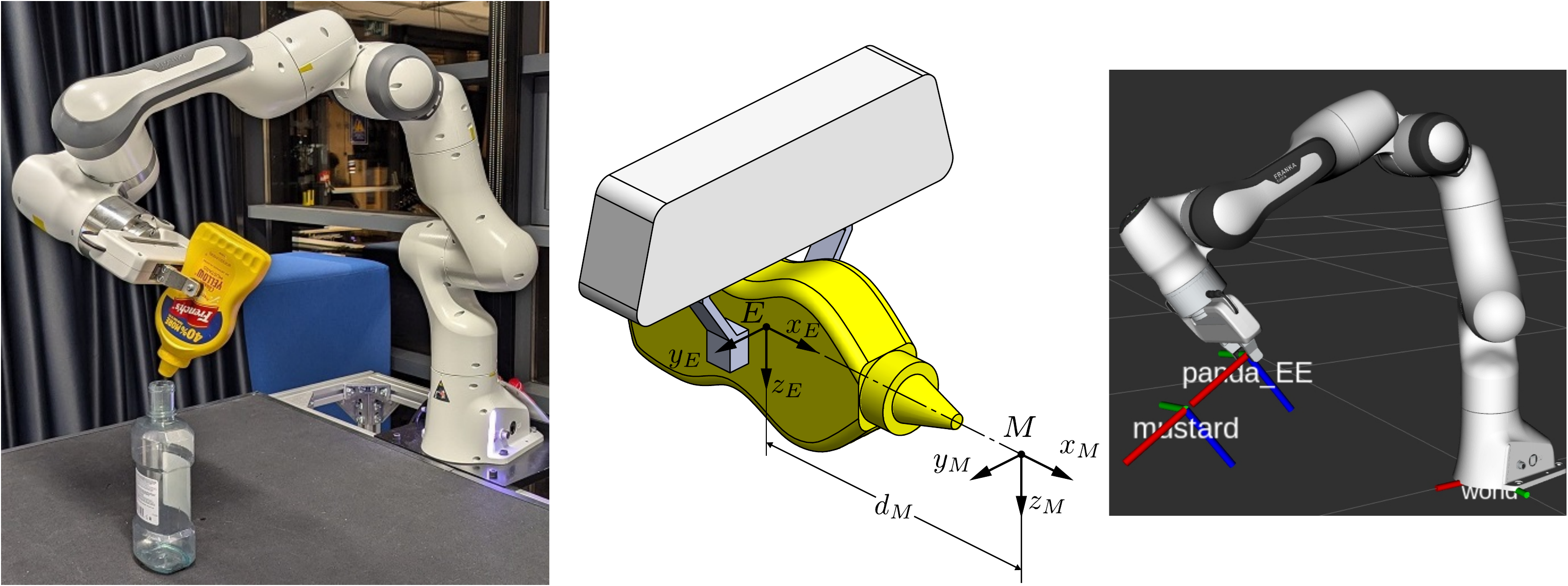}
    \caption{Setup for experiment 1, showing coordinate systems $E$ and $M$.}
    \label{fig:listerine_setup}
\end{figure}

A Franka Emika Panda arm is used to carry out kinesthetic demonstrations of a pouring task as shown in Fig. \ref{fig:listerine_setup}. We would like to fit a probability distribution to the orientations of the mustard bottle being held by the end-effector at the instant of approach. Afterwards, we would like to sample some orientations from such distribution and let the robot plan to reach such orientations. The aim of this random sampling is to imitate a task learnt from human demonstrations in a probabilistic way, reflecting the variability of the demonstrations.

Note that the originally recorded dataset is $\mathscr{D}=\left\{{}^{O}_{E}\mathbf{T}_i\in\mathrm{SE}(3)\right\}_{i=1}^{N}$, where $E$ is the end-effector frame (\texttt{panda\textunderscore EE} in \texttt{LibFranka}). In order to exclusively work with orientations, it can be considered that the mustard bottle is always rotating about a pivotal point that is approximately fixed across all poses in the data. Then we define a frame $M$ with origin at such point and with axes parallel to frame $E$ as shown in Fig. \ref{fig:listerine_setup}. The origin of $M$ is a distance $d_M$ away from the origin of frame $E$ along the $X_E-$ axis. Using least squares with the position part of $\mathscr{D}$, it is possible to find the value of $d_M$ that best approximates the origin of frame $M$ to a fixed point. Hence, for the statistical modelling in this experiment, the data will be solely orientational, i.e. $\mathscr{D}'=\left\{{}^{O}_{M}\mathbf{R}_i\in\mathrm{SO}(3)\right\}_{i=1}^{N}$. 

\begin{figure}[ht]
    \centering
    \includegraphics[width=0.7\textwidth]{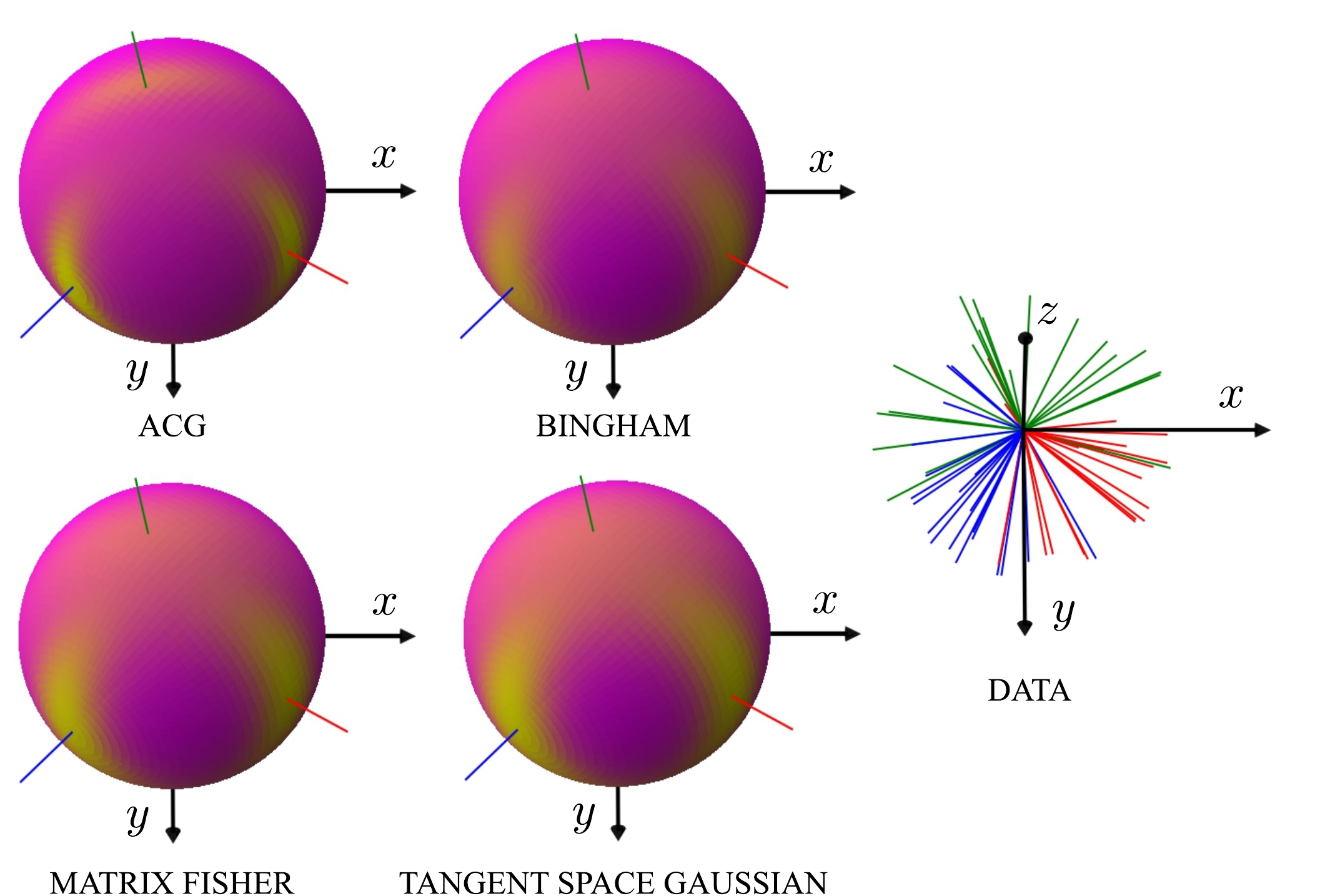}
    \caption{Visualisation of the different models fitted in experiment 1.}
    \label{fig:listerine_distributions}
\end{figure}

\begin{figure}[ht]
    \centering
    \includegraphics[width=1.0\textwidth]{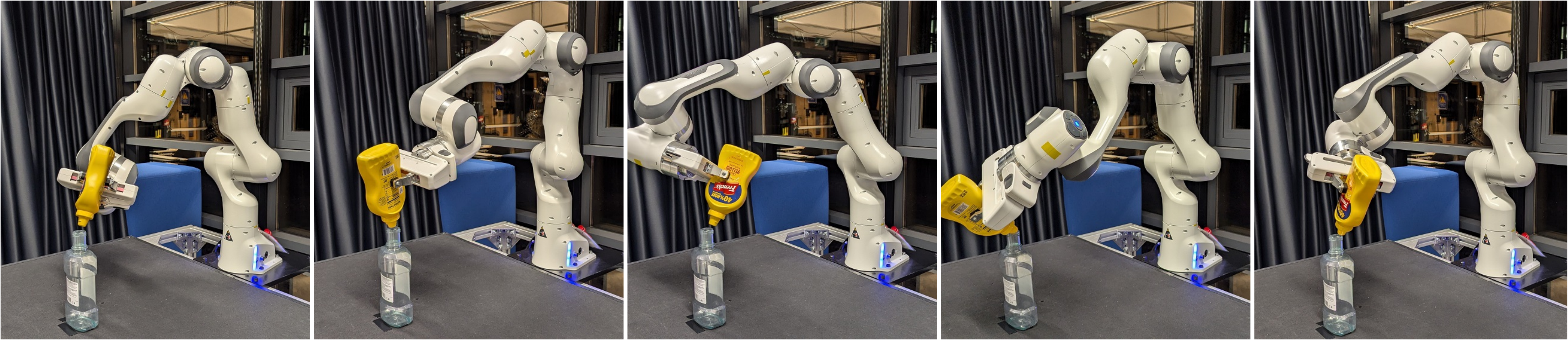}
    \caption{Samples from the fitted ACG distribution. In counterclock-wise order starting from the top-left corner: $(-0.325, 0.441, 0.290, 0.785)^{\top}$, $(0.141, -0.777, -0.009, -0.613)^{\top}$, $(-0.767, 0.229, 0.467, 0.376)^{\top}$, $(0.582, -0.613, -0.241, -0.476)^{\top}$, and $(0.076, 0.893, -0.316, 0.313)^{\top}$, in convention $(x, y, z, w)$.}
    \label{fig:listerine_samples}
\end{figure}

$N=20$ data points were recorded. Those are shown table \ref{tab:data_listerine} in the Appendix, where only quaternion notation is used to save space. The results of estimating the parameters for ACG, Bingham, Matrix Fisher and Gaussian in the tangent space are shown in Table \ref{tab:exp_1}. The visualisation of each of these distribuitons is shown in Fig. \ref{fig:listerine_distributions}. Fig. \ref{fig:listerine_samples} shows the robot taking the mustard bottle to 5 orientations sampled from the fitted ACG distribution.

\begin{table}[h]
    \centering
    \begin{tabular}{cc}
    \hline
    \textbf{Model} & \textbf{Parameters} \\ \hline
    \parbox[c]{3cm}{ACG} &
    \parbox[c][2cm][c]{8cm}{$\hat{\mathbf{\Lambda}}=\left[\begin{array}{cccc} 2.879 & -0.020 & -1.098 & -0.531 \\
     -0.020 & 0.423 & 0.014 & 0.189 \\
     -1.098 & 0.014 & 0.449 & 0.200 \\
     -0.531 & 0.189 & 0.200 & 0.25\end{array}\right]$
     } \\ 
    \hline
    \parbox[c]{3cm}{Bingham} & \parbox[c][2.5cm][c]{8cm}{$\hat{\kappa}_1 = -52.556$, $\hat{\kappa}_2= -20.25$ $\hat{\kappa}_3=-3.24$\\
    $\hat{\mathbf{v}}_1=(-0.362, 0.028, -0.931, -0.026)^{\top}$\\
    $\hat{\mathbf{v}}_2=(-0.137,0.417,0.091,-0.894)^{\top}$\\
    $\hat{\mathbf{v}}_3=(0.084,0.909,-0.017,0.409 )^{\top}$\\
    $\hat{\mathbf{v}}_4=(0.918, -0.010, -0.352, -0.181)^{\top}$
    } \\ 
    \hline
    \parbox[c]{3cm}{Matrix Fisher} & \parbox[c][3.5cm][c]{8cm}{$\hat{\mathbf{U}} = \begin{bmatrix}
    -0.070 & 0.992 & 0.109 \\
    -0.174 & -0.119 & 0.977 \\
    0.982 & 0.050 & 0.181
    \end{bmatrix}$\\
    $\hat{\mathbf{V}}=\begin{bmatrix}
    -0.711 & 0.700 & 0.071 \\
    -0.146 & -0.048 & -0.988 \\
    -0.688 & -0.713 & 0.137
    \end{bmatrix}$\\
    $\hat{\psi}_1=15.866$, $\hat{\psi}_2=8.245$, $\hat{\psi}_3=-6.564$
    } \\ 
    \hline
    \parbox[c]{3.5cm}{Gaussian in $T_{\bm{\upmu}}\mathcal{S}^3$} & \parbox[c][2.5cm][c]{8cm}{
    $\hat{\bm{\upmu}}=(-0.918, 0.010, 0.352, 0.181)^{\top}$\\
    $\hat{\mathbf{D}}=\mathrm{diag}(0.227, 0.033, 0.012)$\\
    $\hat{\mathbf{B}}=\begin{bmatrix}
    0.081 & 0.914 & -0.018 & 0.397 \\
    -0.134 & 0.404 & 0.102 & -0.899 \\
    -0.364 & 0.030 & -0.930 & -0.038
    \end{bmatrix}$
    } \\ 
    \hline
\end{tabular}
    \caption{Parameters estimated for the different models fitting the data in experiment 1.}
    \label{tab:exp_1}
\end{table}

\subsection{Experiment 2: Precision of extrinsic calibration for an RGBD camera}\label{sec:exp_aruco}

\begin{figure}[ht]
    \centering
    \includegraphics[width=0.8\textwidth]{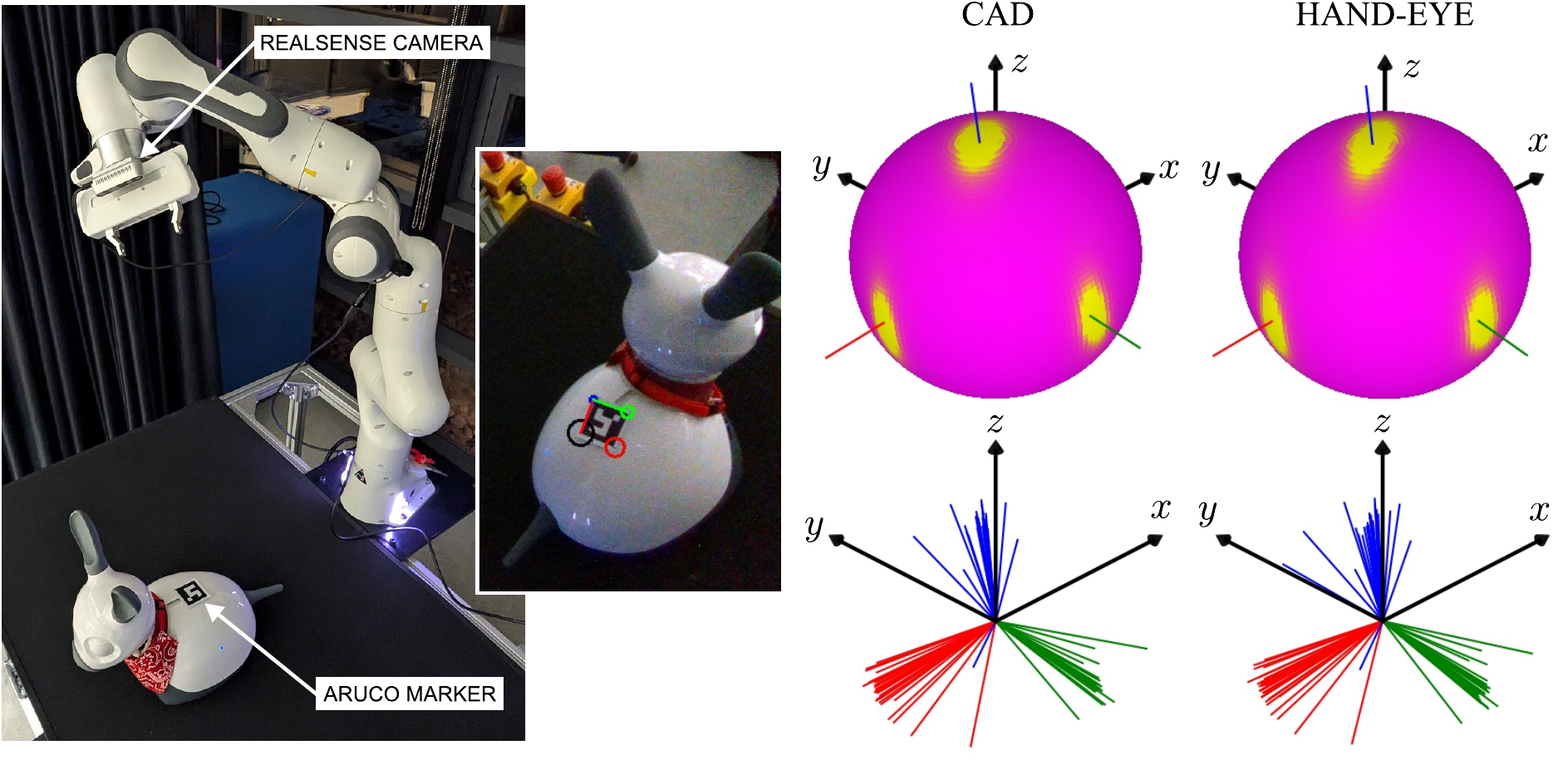}
    \caption{LEFT: Setup for experiment 2, showing the frame $A$ on the back of the Miro-E being detected by the RealSense camera attached to the Franka arm. RIGHT-TOP: Comparison of visualisation of the fitted ACG distribution to $\mathscr{D}_{\mathrm{CAD}}$ and $\mathscr{D}_{\mathrm{HEC}}$. RIGHT-BOTTOM: The datasets $\mathscr{D}_{\mathrm{CAD}}$ and $\mathscr{D}_{\mathrm{HEC}}$}
    \label{fig:aruco}
\end{figure}

Consider a Franka Emika Panda arm with an RGBD RealSense camera attached to its end-effector as shown in Fig. \ref{fig:aruco}. The camera is tracking a frame $A$ defined on an ArUco marker fixed to the back of a Miro-E robot. While the Miro-E is static, the orientation of frame $A$ with respect to the world frame $O$, ${}^{O}_{A}\mathbf{R}$, is read from $N$ different poses of the end-effector. With a perfect calibration (and many other conditions), all those readings should be identical. However, this is never the case in a real experimental setup. In this experiment, we tried two different methods for extrinsic calibration: {\it hand-eye calibration} (HEC) \cite{camera_calibration} and ideal extrinsics using the CAD drawings of the camera and its mounting. It is desired to know which of these two calibration methods gives better precision in orientation. For this sake, and ACG, Bingham, Matrix Fisher and Gaussian in the tangent space were fitted to datasets collected using each calibration method, and the resulting concentration parameters were analysed.

The HEC was performed prior to the experiment following a classical approach \cite{camera_calibration}. An ArUco marker was fixed in a different pose to the one in the subsequent experiment, and the relevant transformation matrices were recorded from 20 different configurations of the robot arm. Hence, HEC was performed with 10 randomised pairs for the calibration equation $\mathbf{A}_i\mathbf{X}=\mathbf{X}\mathbf{B}_i$.

To compare both calibrations, two new datasets were generated: $\mathscr{D}_{\mathrm{HEC}}=\{{}^{O}_{A}\mathbf{R}_{\mathrm{HEC},i}\in\mathrm{SO}(3)\}_{i=1}^N$, which uses HEC, and $\mathscr{D}_{\mathrm{CAD}}=\{{}^{O}_{A}\mathbf{R}_{\mathrm{CAD},i}\in\mathrm{SO}(3)\}_{i=1}^N$ which uses CAD. Each pair $\{{}^{O}_{A}\mathbf{R}_{\mathrm{HEC},i}$, ${}^{O}_{A}\mathbf{R}_{\mathrm{CAD},i}\}$ is obtained from the same $i$th pose of the end-effector. $N=30$ poses were recorded. Due to space constraints, the dataset is not included in this paper, but can be accessed in the repository shared with this paper \cite{rotstats}. 

Table \ref{tab:exp_2} shows the parameters estimated to both datasets. For the ACG, we can have a sense of concentration by looking at the ratio of the first to the second eigenvalues of $\mathbf{\Lambda}$ \--- the larger $a_1/a_2$, the more concentrated the distribution. It can be seen from table \ref{tab:exp_2} that, in the ACG model, $\mathscr{D}_{\mathrm{CAD}}$ is more concentrated. For the Bingham distribution, the larger the {\it magnitude} of the parameters $\kappa_i$, the more concentrated the distribution. Then, table \ref{tab:exp_2} suggests that for the Bingham model, $\mathscr{D}_{\mathrm{CAD}}$ is more concentrated. In the case of the Matrix Fisher distribution, the larger the sum $\psi_j+\psi_k$, the more concentrated the distribution around the $i$th axis of the mean. Then, from \ref{tab:exp_2} the Matrix Fisher model suggests that the concentration is larger around the three axes of the mean for $\mathscr{D}_{\mathrm{CAD}}$. Finally, for the Gaussian in the tangent space of $\mathcal{S}^3$, $\mathscr{D}_{\mathrm{CAD}}$ gets smaller eigenvalues in the covariance matrix of its MVN. Hence, such model also supports the conclusion that $\mathscr{D}_{\mathrm{CAD}}$ is more concentrated.

The right-hand side of Fig. \ref{fig:aruco} shows the visualisation of the fitted ACG distributions. It can be seen that the one corresponding $\mathscr{D}_{\mathrm{CAD}}$ is slightly more concentrated than that for $\mathscr{D}_{\mathrm{HEC}}$. All four models agree with this conclusion. In practice and empirically, we have had better results using CAD than algorithms to calibrate when performing table-top grasping operations in our laboratory. A potential explanation is the proximity of objects to the camera, which is close to the detection minimum. It is expected that an algorithmic calibration can perform better than a CAD one for longer distances. A systematic benchmarking of extrinsic camera calibration methods was presented in \cite{camera_calibration_benchmark}.

\begin{table}[ht]
    \centering
    \begin{tabular}{ccc}
    \toprule
    \textbf{Model} & \textbf{For $\mathscr{D}_{\mathrm{CAD}}$} & \textbf{For $\mathscr{D}_{\mathrm{HEC}}$} \\ \hline
    \parbox[c]{3.5cm}{\centering ACG} &
    \parbox[c][2cm][c]{4cm}{
    $\hat{a}_1=1$, \\
    $\hat{a}_2=1.763\times10^{-3}$, \\
    $\hat{a}_3=4.38\times10^{-4}$, \\
    $\hat{a}_4=1.64\times10^{-4}$}
    &
    \parbox[c][2cm][c]{4cm}{
    $\hat{a}_1=1$,\\
    $\hat{a}_2=2.269\times10^{-3}$,\\ 
    $\hat{a}_3=3.91\times10^{-4}$,\\ 
    $\hat{a}_4=2.43\times10^{-4}$
     } \\ 
    \hline
    \parbox[c]{3.5cm}{\centering Bingham} & 
    \parbox[c][2cm][c]{4cm}{
    $\hat{\kappa}_1=-351.562$, \\
    $\hat{\kappa}_2=-100$,  \\
    $\hat{\kappa}_3=-30.25$,\\ 
    $\hat{\kappa}_4=0$}
    &
    \parbox[c][2cm][c]{4cm}{
    $\hat{\kappa}_1=-333.063$, \\
    $\hat{\kappa}_2=-90.25$,  \\
    $\hat{\kappa}_3=-30.25$, \\
    $\hat{\kappa}_4=0$
    } \\ 
    \hline
    \parbox[c]{3.5cm}{\centering Matrix Fisher} & 
    \parbox[c][2cm][c]{4cm}{
    $\hat{\psi}_1=106.233$, \\
    $\hat{\psi}_2=70.547$, \\
    $\hat{\psi}_3=-55.581$}
    &
    \parbox[c][2cm][c]{4cm}{
    $\hat{\psi}_1=99.212$, \\
    $\hat{\psi}_2=67.418$, \\
    $\hat{\psi}_3=-52.559$
    } \\ 
    \hline
    \parbox[c]{3.5cm}{\centering Gaussian in $T_{\bm{\upmu}}\mathcal{S}^3$} & 
    \parbox[c][1cm][c]{5.5cm}{
    $\hat{\mathbf{D}}=\mathrm{diag}(0.019, 0.005, 0.002)$}
    &
    \parbox[c][1cm][c]{5.5cm}{
    $\hat{\mathbf{D}}=\mathrm{diag}(0.019, 0.006, 0.002)$} 
    \\ 
    \bottomrule
\end{tabular}
    \caption{Concentration parameters estimated for the different models from the two datasets in experiment 2. For the ACG distribution, $\hat{a}_i$ are the eigenvalues of $\hat{\mathbf{\Lambda}}$}
    \label{tab:exp_2}
\end{table}

\section{Conclusions}\label{sec:conclusions}

\noindent A substantial number of models for distributions of orientational data were summarised in this paper. The information was presented in a concise way to only offer the necessary information to calculate density, estimate parameters and simulate. 

For the 1-DOF orientational data, it can be concluded that the von Mises distribution has a simpler density function than the wrapped normal. However, the latter is easier to simulate. In the case of models for 2-DOF orientational data, it was seen that the simplest case is the von Mises Fisher. Nevertheless, it has the drawback of being isotropic. Parameter estimation from data is also difficult for the von Mises Fisher distribution. On the other hand, the Kent and the ESAG distributions are anisotropic and are similar to each other. However, the latter is much simpler to simulate and has a simpler, closed-form normalising constant. 

The advantages of the ESAG over the Kent distribution are a consequence of the former being part of the family of angular Gaussian distributions while the latter coming from the Bingham family. This fact also makes the ACG distribution more convenient for 3-DOF orientational data. While both the Bingham and the matrix Fisher models have awkward and computationally expensive normalising constants and MLEs, these properties are easy to compute for the ACG. In addition, sampling from both Bingham and matrix Fisher distributions is done using an ACG envelop. 

The definition of Gaussians in the tangent space of spheres and in the Lie algebra of SO(3) was also discussed. It was seen that such models can be computationally cheap and simple but are only recommended for concentrated data.

Diffusion equations and their solutions were briefly discussed. Those models are less restrictive, allowing to break symmetries, for example, by the introduction of a drift term. This comes at the cost of PDFs in terms of infinite summations and potential complexity to solve the desired diffusion equation.

We also applied some of those models to two cases of real data. From such experiments and the theory reviewed in this paper, it can be concluded that the ACG distribution is the most convenient model. Unfortunately, not much attention has been given to this model in engineering and computer science, where the Bingham and Gaussians in the tangent space are the most commonly used. 

It is also important to mention that in many applications the data are in SE(3), i.e. they are poses with translational and orientational components. In many cases, it is possible to decouple these two components. The position is then analysed using the MVN distribution, and for the orientation we use the models presented here. Any coupling in SE(3) data must be done cautiously due to the lack of a bi-invariant metric in the manifold, which may lead to units-dependent results.

\section*{Acknowledgment}

\noindent The author thanks the Cobot Maker Space of the University of Nottingham, especially Dominic Price, for kindly allowing him to use their equipment for the experiments reported in this paper. The author also thanks Prof. Simon Preston, from the University of Nottingham for his valuable input. Finally, the author also thanks one of the anonymous reviewers who suggested the inclusion of models from diffusion equations for completeness of this survey.


\bibliographystyle{siam}
\bibliography{orientations}  

\appendix
\section{Relevant expressions}\label{sec:appendix_expressions}

The modified Bessel function of the first kind and order $v$ are defined as follows:

\begin{equation}\label{eq:Iv}
    I_{\nu}(z) =  \sum^{\infty}_{n=0}\frac{1}{\Gamma(\nu+n+1)n!}\left(z\right)^{2n+\nu}
\end{equation}

\noindent where $\Gamma(\cdot)$ is the the gamma function. In the cases of $\nu=0$ and $\nu=1$, (\ref{eq:Iv}) reduces to, respectively:
\begin{equation}\label{eq:I0_I1}
    I_0(z) = \sum^{\infty}_{n=0}\frac{\left(\frac{1}{4}z^2\right)^n}{\left(n!\right)^2},\;\;
    I_1(z) = \frac{z}{2}\sum^{\infty}_{n=0}\frac{\left(\frac{1}{4}z^2\right)^n}{n!(n+1)!}
\end{equation}

\section{Dataset for experiment 1}\label{sec:appendix_listerine}
Table \ref{tab:data_listerine} shows the elements of the dataset $\mathscr{D}'$ in Section \ref{sec:exp_listerine}.
\begin{table}[]
    \centering
    \begingroup
    \footnotesize
    \begin{tabular}{|cccc|cccc|}
    \hline
    $x$ & $y$ & $z$ & $w$ & $x$ & $y$ & $z$ & $w$ \\ \hline
    0.916 & -0.084 & -0.282 & -0.274 &
    0.772 & -0.362 & -0.425 & -0.304 \\ 
    0.968 & -0.101 & -0.175 & -0.146 &
    0.884 & 0.078 & -0.304 & -0.346 \\ 
    -0.634 & 0.087 & 0.364 & 0.677 &
    -0.595 & 0.744 & 0.226 & 0.205 \\ 
    0.731 & 0.474 & -0.448 & 0.200 &
    0.937 & 0.301 & -0.135 & -0.117 \\ 
    0.747 & -0.366 & -0.333 & -0.444 &
    -0.504 & 0.722 & 0.171 & 0.443 \\ 
    0.641 & 0.750 & -0.147 & -0.068 &
    0.822 & 0.412 & -0.366 & 0.140 \\ 
    0.891 & -0.225 & -0.395 & 0.003 &
    0.703 & -0.436 & -0.367 & -0.426 \\
    0.910 & -0.064 & -0.347 & -0.219 &
    0.840 & 0.371 & -0.394 & -0.042 \\ 
    0.979 & 0.027 & -0.200 & -0.009 &
    0.925 & 0.205 & -0.317 & -0.044 \\ 
    0.741 & 0.468 & -0.431 & 0.215 &
    0.907 & -0.149 & -0.389 & -0.053 \\ \hline
    \end{tabular}
    \endgroup
    \caption{Quaternions in the dataset for experiment 1, Section \ref{sec:exp_listerine}}
    \label{tab:data_listerine}
\end{table}

\end{document}